\documentclass{article}
\usepackage[utf8]{inputenc}
\usepackage{amsmath}
\usepackage{geometry}
\usepackage{times}
\usepackage{graphicx}
\usepackage{hyperref}
\usepackage[numbers]{natbib}
\usepackage{caption}
\usepackage{float}
\usepackage[T1]{fontenc}
\usepackage{amssymb}
\usepackage{booktabs}
\usepackage{subcaption}
\usepackage{authblk}

\title{ENERGY-DEPENDENT ANISOTROPY OF COSMIC-RAY
MUONS: A TWELVE-YEAR STUDY WITH ICECUBE
NEUTRINO OBSERVATORY}

\author[1]{Nabin Upadhya Dhakal}
\author[1]{Nabin Bhusal}
\author[1]{Manjeet Kunwar\thanks{Corresponding author: manjeetkunwar04@gmail.com}}
\author[1]{Manil Khatiwada}
\author[1]{Shiv Narayan Yadav}

\affil[1]{Central Department of Physics, Tribhuvan University, Kirtipur, Kathmandu 44618, Bagmati, Nepal}
    
\date{}

\begin{document}

	\maketitle
    
	\begin{abstract}
\noindent
Understanding the anisotropic behaviour of cosmic rays is vital for uncovering their origin, propagation, and interaction with galactic magnetic fields. Although previous studies have reported large-scale anisotropy in cosmic-ray arrival directions, the energy dependence of such features remains incompletely resolved, especially in the Southern Hemisphere. In this work, we present a comprehensive, energy-resolved analysis of cosmic-ray muon anisotropy using twelve years of data (2011–2023) from the IceCube Neutrino Observatory. The dataset comprises $7.92\times10^{11}$ reconstructed muon events, spanning a reconstructed energy range of $\sim13$~TeV to 5.3~PeV, binned into nine logarithmic intervals. We divide the spectrum at 5~GeV (in log-scaled units) to contrast low- and high-energy anisotropy features. Our methodology combines sidereal modulation analysis, angular distribution profiling, Fourier power spectrum evaluation, sidereal-angular heatmap visualisation, and full-sky HEALPix intensity mapping. Additionally, we apply Gaussian and power-law fitting to reconstructed energy distributions and evaluate the model fidelity using $\chi^2$, reduced $\chi^2_\nu$, and Bayesian Information Criterion (BIC). The results reveal a clear energy-dependent evolution in anisotropy. Low-energy muons ($E \leq 5$~GeV) exhibit strong dipolar modulation, elevated harmonic content, and pronounced large-scale structures consistent with geomagnetic and atmospheric influences. In contrast, high-energy muons ($E > 5$~GeV) display subtler, localised features, diminished modulation amplitude, and greater directional coherence, suggesting reduced magnetic scattering and increasing dominance of source-related effects. The energy distributions show excellent Gaussian behaviour in mid and high energy bins, with the 6.50–100 bin yielding the lowest BIC and $\chi^2_\nu$, validating IceCube’s energy reconstruction pipeline even at PeV scales. These findings offer robust, long-term confirmation of energy-dependent anisotropy in cosmic-ray muon arrival directions. The observed transition from environmental modulation to astrophysical structure reflects key predictions from cosmic-ray diffusion theory. This study not only enhances IceCube’s role as a precision cosmic-ray observatory but also establishes a statistical and observational framework for future anisotropy research in the PeV regime.\\

\\

	\end{abstract}

	\section{Introduction}

Cosmic rays are highly energetic charged particles, predominantly protons and heavier nuclei, that traverse interstellar and intergalactic space before reaching the Earth’s atmosphere\cite{Gaisser1990cosmic}. Upon interaction with atmospheric nuclei, these primary cosmic rays generate extensive air showers composed of secondary particles, including muons, electrons, and neutrinos\cite{Gaisser2012,Honda2004}. Among these, muons play a particularly significant role in cosmic-ray studies due to their long lifetimes, deep penetration capabilities, and the partial retention of directional information from the parent primary particles\cite{IceCubeMuon2016,IceCubeResolution2014}. For decades, cosmic rays were assumed to be isotropic at Earth due to the randomising effects of galactic magnetic fields on charged particle trajectories\cite{Jokipii1966diffusion}. However, high-precision measurements from ground-based and underground detectors have repeatedly revealed statistically significant anisotropies in cosmic-ray arrival directions, particularly at TeV to PeV energies\cite{Abbasi2010}. These anisotropies manifest across a variety of angular scales from broad dipole-like modulations\cite{Amenomori2006largeanisotropy,Abbasi2011} to localised excess regions\cite{McNally2021}, offering a valuable probe into the structure of galactic magnetic fields\cite{Blasi2012}, cosmic-ray propagation mechanisms\cite{Ptuskin2006}, and potentially, the location of nearby astrophysical sources\cite{Giacinti2012}. The advent of large-volume neutrino detectors such as the \textit{IceCube Neutrino Observatory}, located at the South Pole\cite{IceCubeReview2017,IceCube2006}, has revolutionised our ability to study high-energy cosmic-ray anisotropies with unprecedented statistical power\cite{Aartsen2016,Aartsen2024}. While IceCube’s primary mission is to detect neutrinos via Cherenkov radiation emitted by secondary charged particles in ice, its sensitivity to the enormous background of atmospheric muons, generated by cosmic-ray interactions, renders it a powerful instrument for mapping cosmic-ray arrival directions in the Southern Hemisphere\cite{IceCubeMuon2016,IceCubeResolution2014}. Recent analyses, including those by IceCube, HAWC, Tibet AS$\gamma$, and Super-Kamiokande, have reported energy-dependent features in cosmic-ray anisotropies across a wide energy spectrum~\cite{IceCube:2023gpr,Abeysekara:2018qho,Amenomori:2005dy,Guillian:2005wp}.

\href{https://arxiv.org/search/astro-ph?searchtype=author&query=Abbasi,+R}{R. Abbasi}, \href{https://arxiv.org/search/astro-ph?searchtype=author&query=Ackermann,+M}{M. Ackermann},  The recent work analysed $7.92 \times 10^{11}$ cosmic-ray–induced muon events from 2011 to 2023, providing the most precise measurement of arrival directions in the Southern Hemisphere. Improved simulations and data handling reduced systematics below statistical uncertainties, confirming an energy-dependent anisotropy structure, notably between 100–300~TeV with small binned energy\cite{abbasi2024observation}.  This high-statistics dataset reveals improved significance of PeV-scale anisotropy down to $6^\circ$ and energy-dependent changes in the angular power spectrum, with large-scale features decreasing above 100~TeV\cite{IceCube:2023gpr}. 

This study presents a systematic, energy-resolved analysis of cosmic-ray muon anisotropy using twelve years of IceCube data (May 2011–May 2023), comprising $\sim7.92 \times 10^{11}$ events across nine logarithmic energy bins from $\sim$13~TeV to 5.3~PeV. Using sidereal modulation analysis, angular profiling, power spectra, heatmaps, and HEALPix sky maps\cite{Gorski2005}, we investigate how anisotropic features evolve with energy. Gaussian and power-law fits to energy distributions further validate the statistical reliability of IceCube’s reconstruction and quantify the strength and spatial structure of the observed anisotropy\cite{Abbasi2010sidereal}. However, much of the focus has been on limited energy intervals or shorter timescales. There remains a compelling need to investigate how the anisotropy evolves over a broad energy spectrum and a longer observational baseline.

\section{Data Reduction and Methodology}

We utilise publicly available data released by the IceCube Collaboration under the title \textit{"Observation of Cosmic-Ray Anisotropy in the Southern Hemisphere with Twelve Years of Data Collected by the IceCube Neutrino Observatory"}~\cite{IceCubeDataReleases2025}. This dataset comprises twelve years of reconstructed cosmic-ray muon events collected by the IceCube in-ice array from May 13, 2011, to May 12, 2023, corresponding to a total livetime of 4,295.29 days. The full sample includes approximately $7.92 \times 10^{11}$ events and covers a reconstructed energy range from $\sim$13~TeV to 5.3~PeV. These data were used to construct sky maps of cosmic-ray arrival directions in J2000 equatorial coordinates, facilitating detailed studies of anisotropy across nearly three orders of magnitude in energy. The event sample is subdivided into nine energy intervals, each characterised by a median energy value: 13~TeV, 24~TeV, 42~TeV, 67~TeV, 130~TeV, 240~TeV, 470~TeV, 1,500~TeV, and 5,300~TeV. Each bin is associated with a corresponding FITS (Flexible Image Transport System) sky map file containing both observed event counts and isotropic reference counts in equatorial coordinates. The filenames reflect the log-scaled energy intervals (in GeV): 13~TeV (IC86\_sid\_4–4.25GeV.fits), 24~TeV (IC86\_sid\_4.25–4.5GeV.fits), 42~TeV (IC86\_sid\_4.5–4.75GeV.fits), 67~TeV (IC86\_sid\_4.75–5GeV.fits), 130~TeV (IC86\_sid\_5–5.25GeV.fits), 240~TeV (IC86\_sid\_5.25–5.5GeV.fits), 470~TeV (IC86\_sid\_5.5–6GeV.fits), 1500~TeV (IC86\_sid\_6–6.5GeV.fits), and 5300~TeV (IC86\_sid\_6.5–100GeV.fits). These files collectively enable an energy-resolved investigation of anisotropy using a consistent sky projection across all bins.

To characterise the energy dependence of cosmic-ray anisotropy, we perform a comprehensive analysis of muon arrival directions across the twelve years. The dataset is divided into low-energy ($\leq 5~\mathrm{GeV}$, log scale) and high-energy ($> 5~\mathrm{GeV}$) regimes, allowing for a direct comparison of anisotropic features. Sidereal modulation analysis of event rates is employed to identify temporal variations indicative of anisotropy, with stronger modulations expected at lower energies due to geomagnetic and atmospheric effects. To investigate angular structure and spectral trends, we analyse the angular distributions of events and compute the corresponding Fourier power spectra. Additionally, two-dimensional sidereal-angular heatmaps are constructed to visualise count-based anisotropy signatures across energy bins. To support the evaluation of IceCube’s energy reconstruction performance, we use a simulated dataset provided in \texttt{energy\_distributions.json}, which contains true particle energy histograms for each of the nine bins. Gaussian fits are applied to the reconstructed energy distributions, and the goodness-of-fit is assessed using statistical measures including chi-square ($\chi^2$), reduced chi-square ($\chi^2_\nu$), and the Bayesian Information Criterion (BIC). Comparative statistical analyses are performed separately for the low- and high-energy datasets to evaluate the consistency, resolution, and reliability of the reconstructed energy framework. Overall, this multi-pronged approach provides a robust framework for probing the energy-dependent evolution of cosmic-ray anisotropy, enabling us to test predictions from cosmic-ray diffusion theory and assess IceCube’s capability as a long-term cosmic-ray observatory.

\section{Results and Discussion}

\subsection{Sidereal Modulation of Event Rates}

To probe time-dependent anisotropies in cosmic-ray muon arrival directions in energy dependency, we analysed IceCube data binned in sidereal time. Events were classified into two primary energy groups: low-energy (\(\leq 5\)~GeV) and high-energy ($>$ 5GeV). For each sidereal bin, the normalised event rate was computed and compared across energy bands, making low- and high-energy groups to understand energy flavour.

\begin{figure}[h]
	\centering
	\begin{minipage}{0.49\linewidth}
		\includegraphics[width=\linewidth]{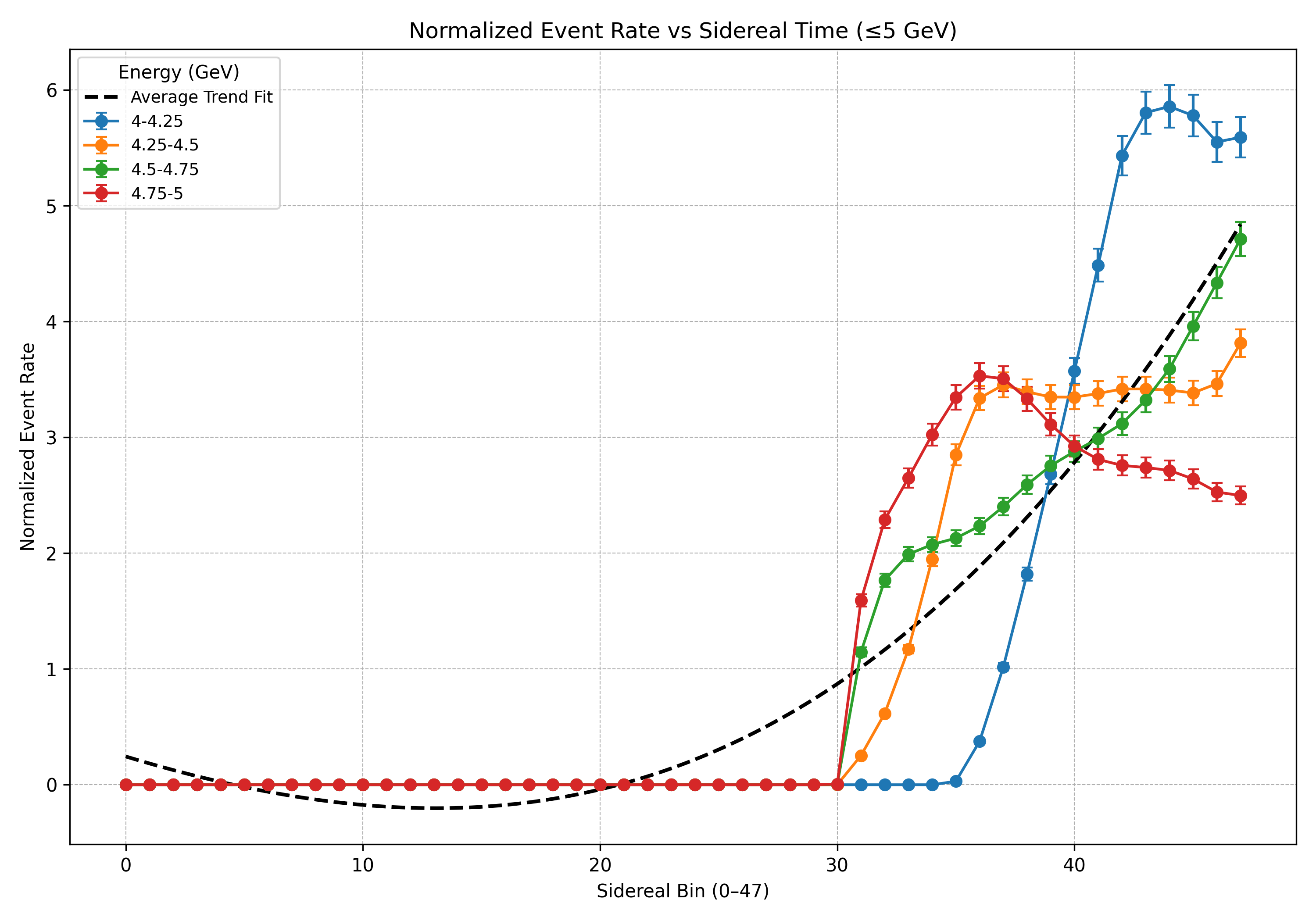}
		\caption{Normalized event rate versus sidereal bin for low-energy events (\(\leq 5\)~GeV). Each curve represents a different energy sub-bin.}
		\label{fig:low_energy_sidereal}
	\end{minipage}
	\hfill
	\begin{minipage}{0.49\linewidth}
		\includegraphics[width=\linewidth]{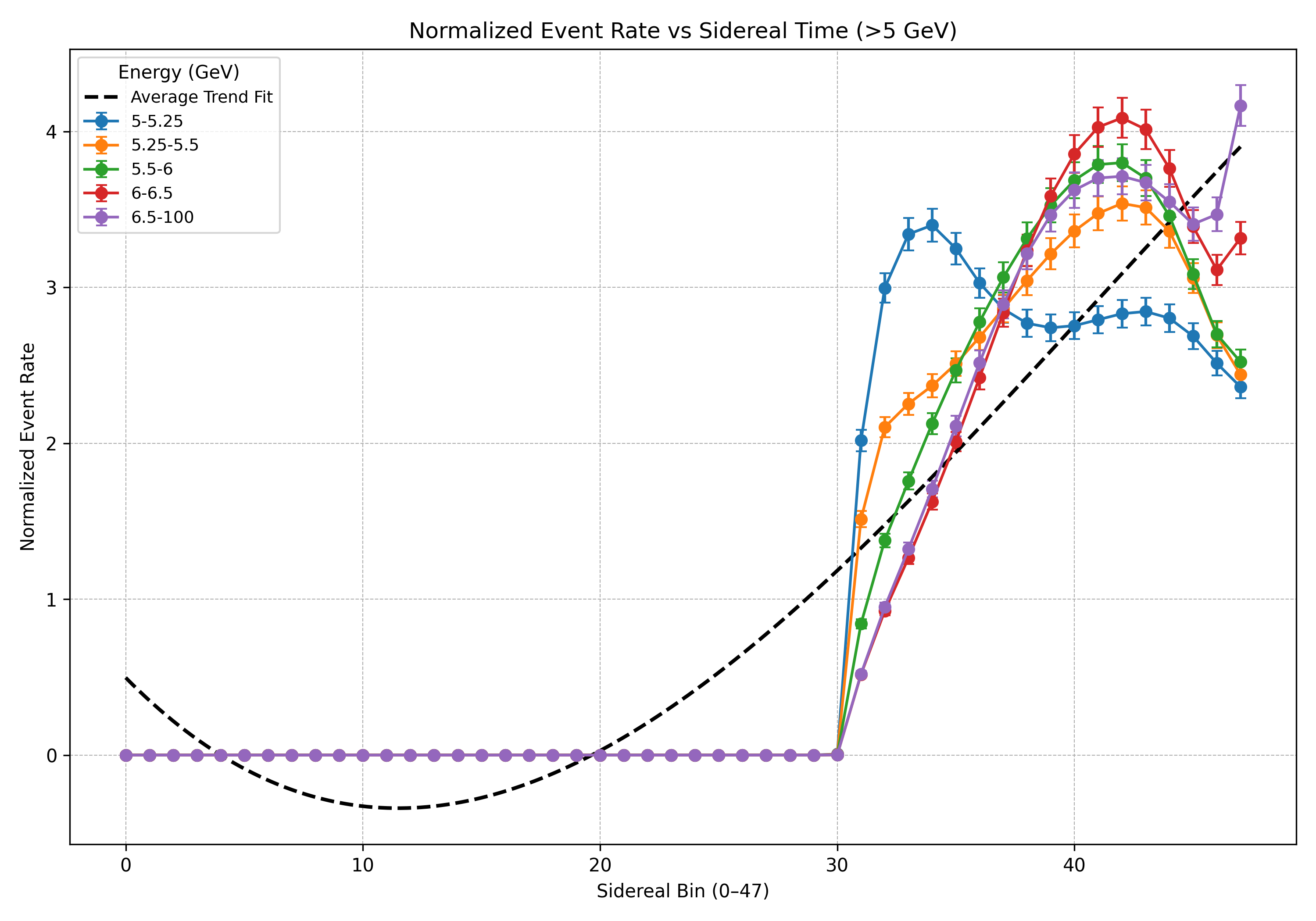}
		\caption{Normalized event rate versus sidereal bin for high-energy events (\(> 5\)~GeV). Each curve represents a different energy sub-bin.}
		\label{fig:high_energy_sidereal}
	\end{minipage}
\end{figure}

Figure~\ref{fig:low_energy_sidereal} shows significant sidereal modulation in the 30–47 bin range for $E \leq 5$\, GeV, indicating strong time-dependent features likely influenced by atmospheric or geomagnetic effects. The modulation amplitude is $\sim$2.08 with a first-harmonic power of 1640.27, highlighting pronounced dipolar anisotropy in low-energy cosmic-ray muons. In contrast, Figure~\ref{fig:high_energy_sidereal} displays subtler but more structured variations in higher-energy bins, with deviations above 6.5, GeV possibly reflecting source-related or detector-induced anisotropies. For $E > 5GeV$, the modulation amplitude is $\sim$1.80 and the first harmonic power is 1624.37, indicating weaker dipolar anisotropy and reduced large-scale structure due to energy-dependent diffusion. Low-energy cosmic rays are more sensitive to the galactic magnetic field and local source structures, and hence show stronger anisotropies.

The sidereal modulation analysis, breaking the energy band from $5 GeV$-  upper and lower energy bands- shows a clear energy-dependent behaviour in cosmic ray anisotropies. Low-energy muons ($E \leq 5$\, GeV) exhibit strong dipolar modulation and significant large-scale anisotropy, likely influenced by atmospheric and geomagnetic effects. In contrast, higher-energy muons ($E > 5$\, GeV) show weaker modulation with more structured but subtler features, suggesting a transition to source-related or propagation-induced anisotropy. The decrease in modulation amplitude and large-scale power at higher energies is consistent with increased cosmic-ray diffusion and loss of directional coherence.

\subsection{Angular Distribution Analysis}
\begin{figure}[h]
	\centering
	\begin{minipage}{0.49\linewidth}
		\includegraphics[width=\linewidth]{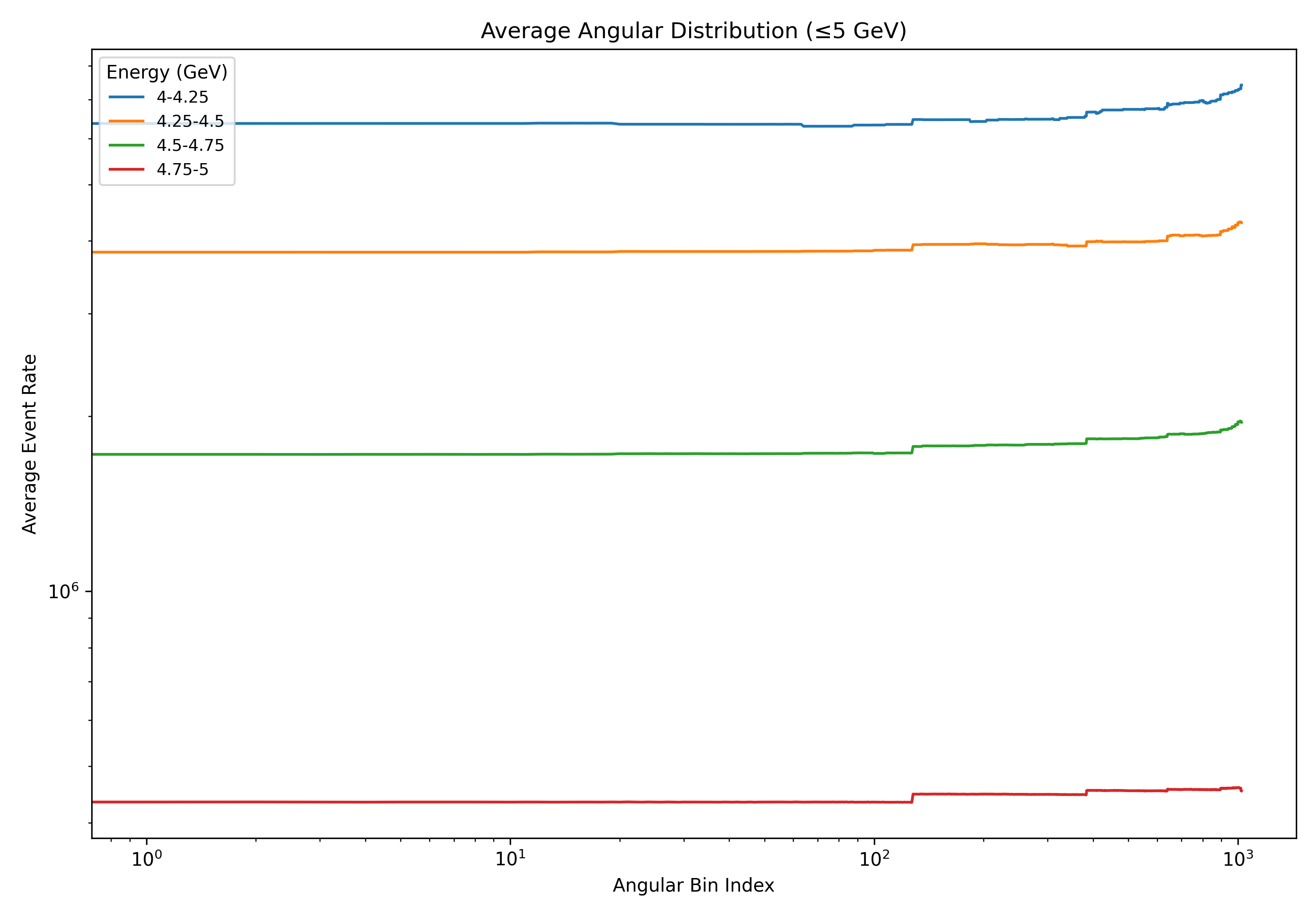}
		\caption{Average angular distribution of event rates for low energy intervals ($\leq 5$~GeV). The x-axis represents the angular bin index, and the y-axis (log scale) shows the average event rate per bin.}
		\label{fig:low-energy}
	\end{minipage}
	\hfill
	\begin{minipage}{0.49\linewidth}
		\includegraphics[width=\linewidth]{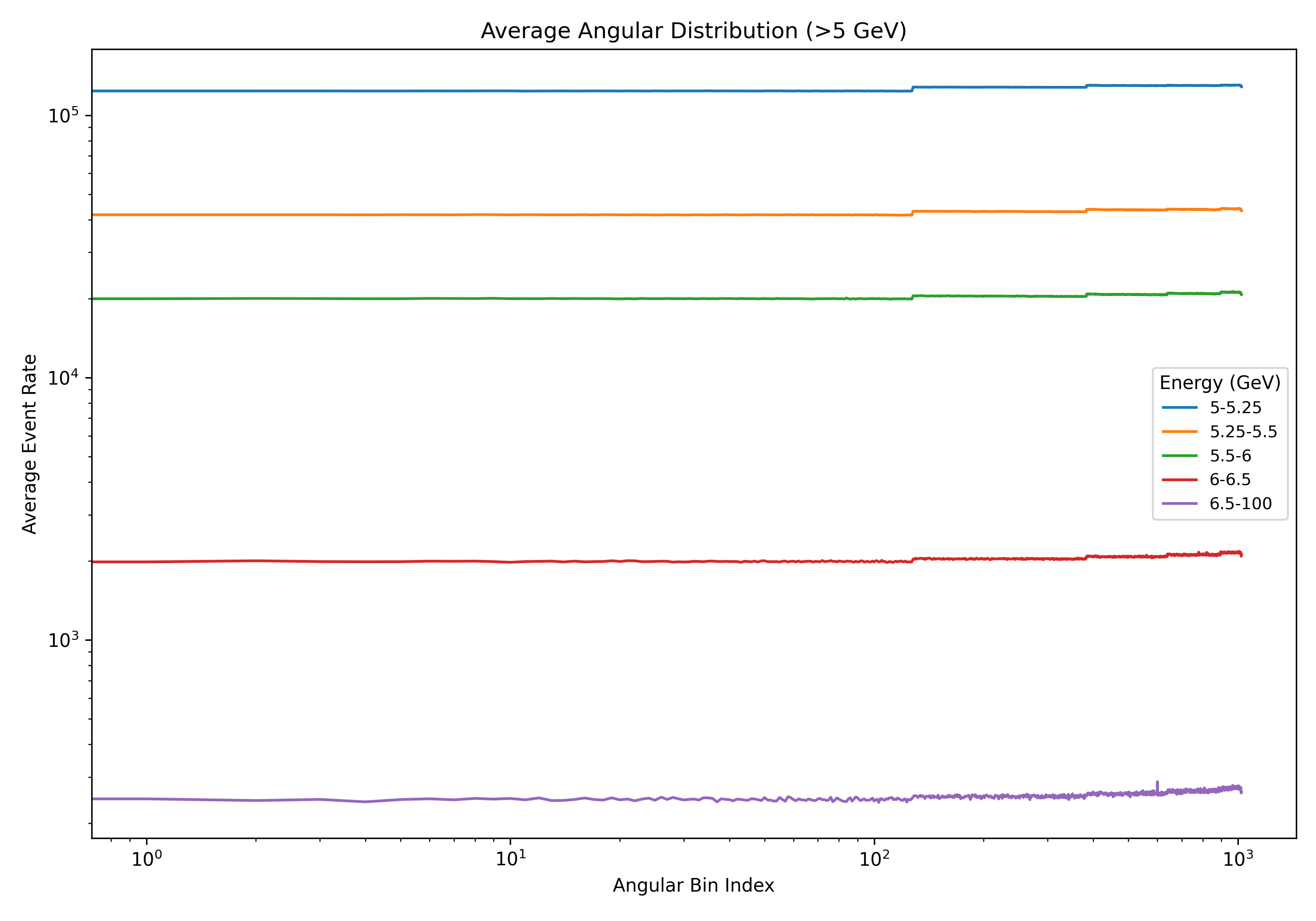}
		\caption{Average angular distribution of event rates for higher energy intervals ($> 5$~GeV). The angular bin index is plotted on the x-axis, and the log-scaled y-axis indicates average event rate.}
		\label{fig:high-energy}
	\end{minipage}
\end{figure}

To investigate the angular distribution of cosmic-ray muon events across various energy intervals, we analysed binned event rate data collected by the IceCube Neutrino Observatory. The analysis was conducted in sidereal time, with the average event rate computed for each angular bin within the defined energy subranges. The resulting angular distributions, presented in Figures~\ref{fig:low-energy} and~\ref{fig:high-energy}, reveal characteristic features of the event topology at both low and high energies, offering insight into the directional behaviour and anisotropic structure of the cosmic-ray flux as a function of energy.

At lower energies ($\leq 5$~GeV), the total event rate exceeds $10^7$ counts per angular bin, exhibiting a broad and relatively flat distribution. As the energy increases, the average event rate decreases by more than two orders of magnitude, consistent with the steeply falling cosmic-ray spectrum. Although the high-energy ($E > 5$\, GeV) angular distributions appear statistically flatter due to reduced counts, subtle structures observed at larger angular indices may suggest the presence of instrumental effects or geomagnetic asymmetries influencing the directional response.

\begin{figure}[h!]
	\centering
	\includegraphics[width=0.8\linewidth]{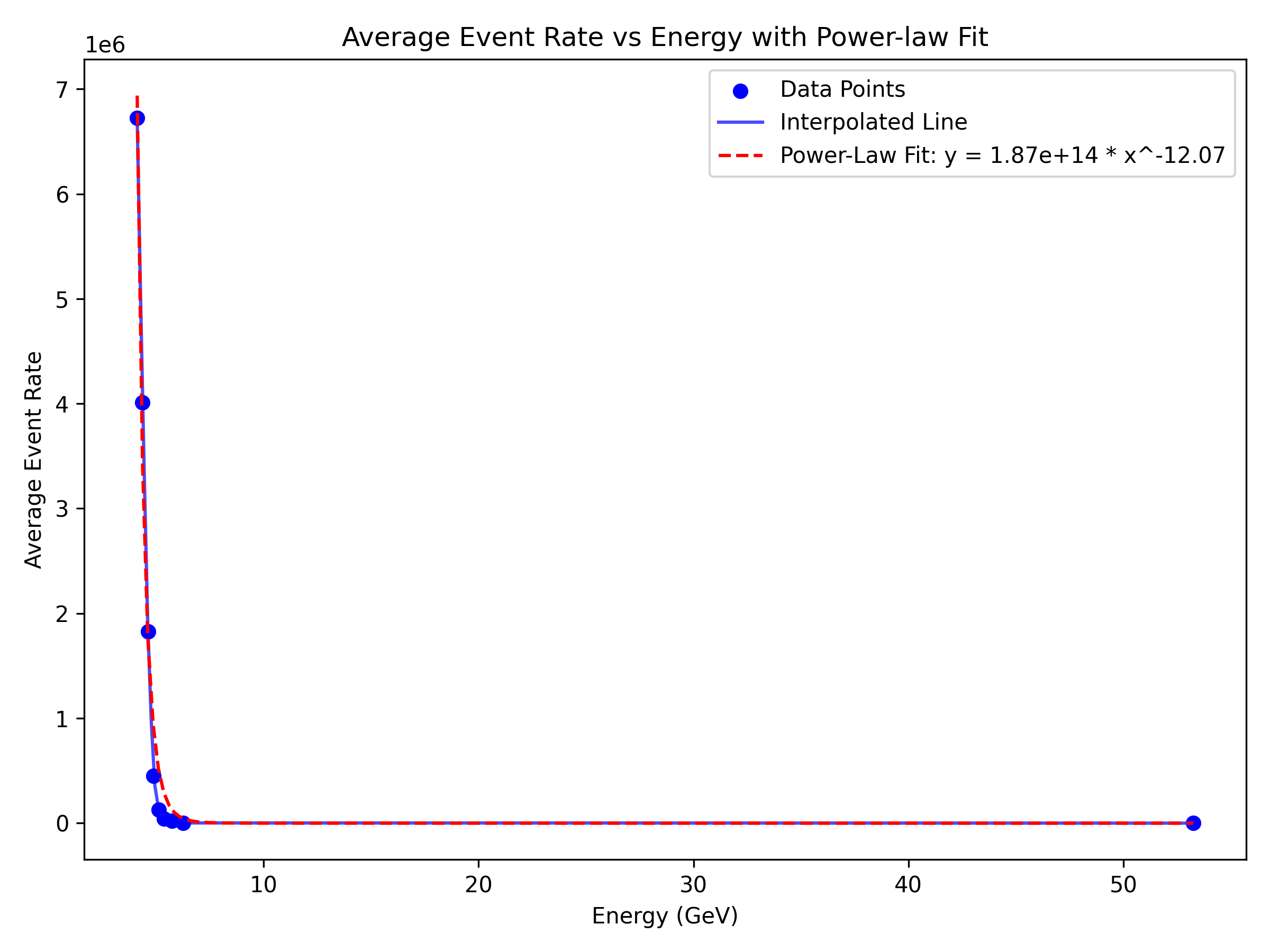}
	\caption{Average event rate vs energy with a power-law fit. The fitted model $y = 1.87 \times 10^{14} \cdot x^{-12.07}$ highlights a steep falloff in muon counts with energy.}

	\label{fig:Power_law}
\end{figure}

The event rate exhibits a sharp decline with increasing energy, consistent with the steeply falling power-law behaviour characteristic of the cosmic-ray spectrum. Figure~\ref{fig:Power_law} illustrates the power-law fit applied to the IceCube data, demonstrating close agreement between the measured data points and the fitted model. A linear interpolation is used to visually connect the discrete data points across energy bins. The best-fit power-law parameters are determined to be $a = 1.868 \times 10^{14}$ and $b = -12.073$, where the functional form of the fit follows $R(E) = a \cdot E^{b}$, with $R(E)$ denoting the event rate and $E$ the reconstructed energy. These results confirm that the energy dependence of the observed muon flux follows the expected cosmic-ray power-law trend across the analysed energy range.

\subsection{ Power Spectrum Analysis}

\begin{figure}[h!]
	\centering
	\begin{minipage}{0.49\linewidth}
		\includegraphics[width=\linewidth]{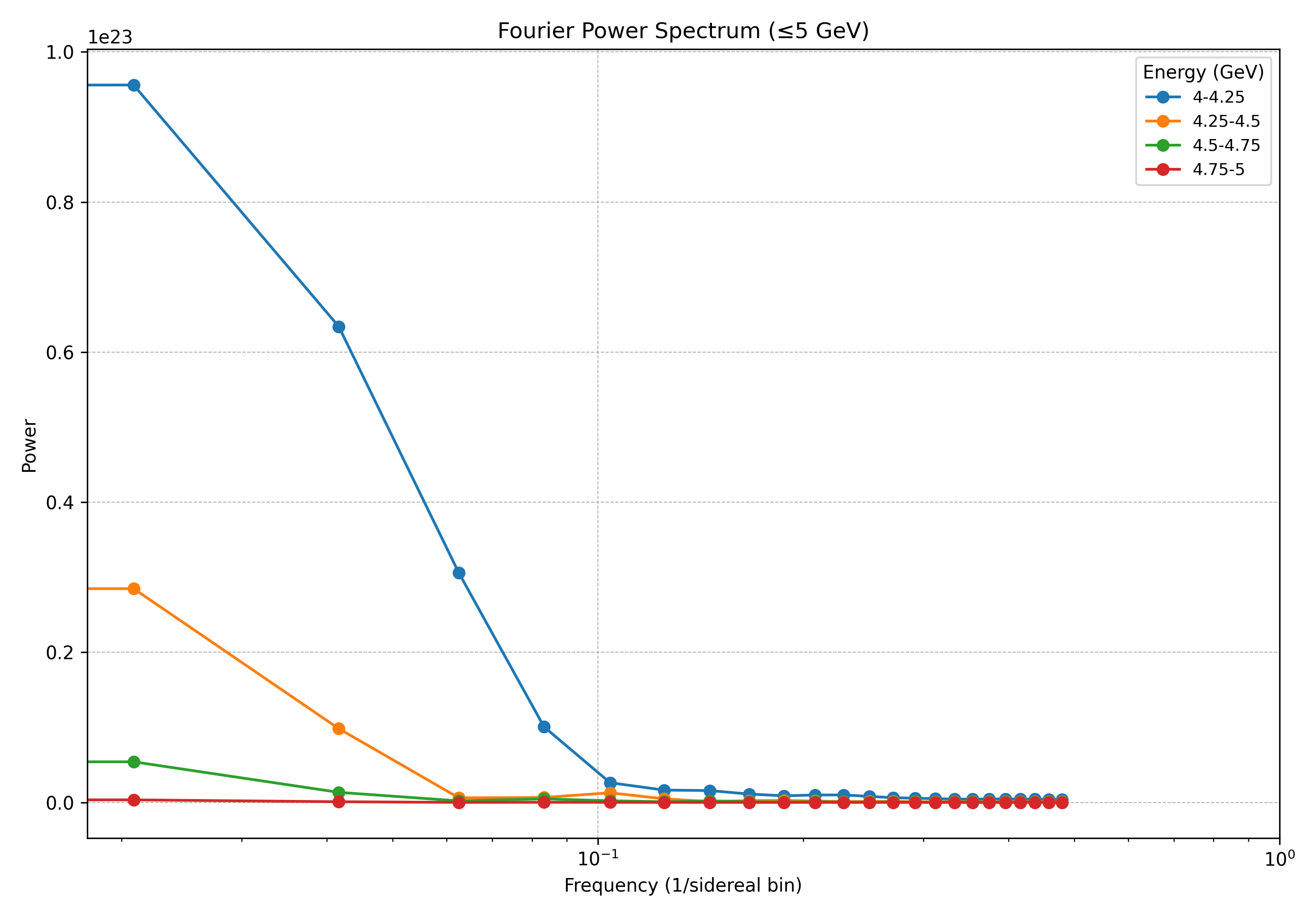}
		\caption{Fourier Power Spectrum for events with energy $\leq$5~GeV. Power is strongest at low frequencies.}
		\label{fig:low_energy}
	\end{minipage}
	\hfill
	\begin{minipage}{0.49\linewidth}
		\includegraphics[width=\linewidth]{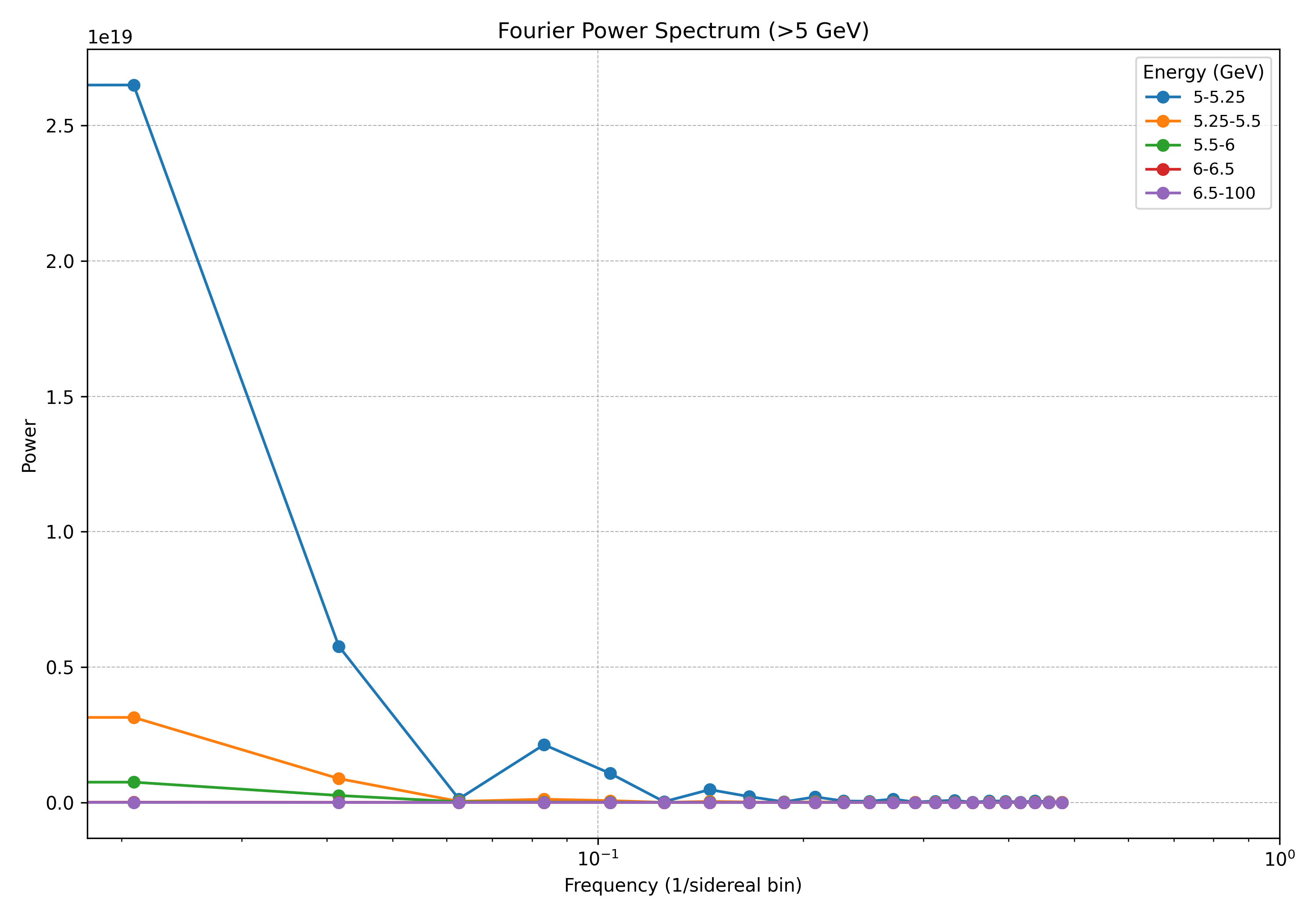}
		\caption{Fourier Power Spectrum for events with energy $>$5~GeV. Power diminishes with increasing energy.}
		\label{fig:high_energy}
	\end{minipage}
\end{figure}

To identify periodic features in the sidereal time signal, we computed the power spectrum of event rate distributions for each energy bin. The resulting spectra, shown in Figures~\ref{fig:low_energy} and~\ref{fig:high_energy}, reveal harmonic components that offer insight into long-timescale modulations in the cosmic-ray flux. These spectral features help quantify the presence and strength of anisotropy across different energy regimes.

A comparative analysis of the Fourier power spectra below and above 5\, GeV reveals a clear energy dependence in the large-scale anisotropy. For muons with $ E\leq5$ \, GeV, the power spectrum exhibits a strong signal at low frequencies, particularly in the 4–4.25\, GeV bin, with peak power reaching $\sim10^{23}$. The power decreases rapidly with increasing frequency, indicating a dominant dipolar or large-scale component. In contrast, for $E > 5$\, GeV, the overall power is significantly lower (maximum $\sim10^{19}$), and the spectral features flatten more quickly with energy. The highest-energy bin (6.5–100\, GeV) shows minimal power across all frequencies. This trend suggests that the anisotropy amplitude diminishes with energy, consistent with increased magnetic diffusion and reduced coherence of cosmic-ray arrival directions at higher energies. The suppression of large-scale features at higher energies aligns with the expected de correlation of arrival directions due to increased rigidity.

\subsection{Sidereal-Angular Heatmap Visualization}

\begin{figure}[H]
	\centering
	\includegraphics[width=0.49\textwidth]{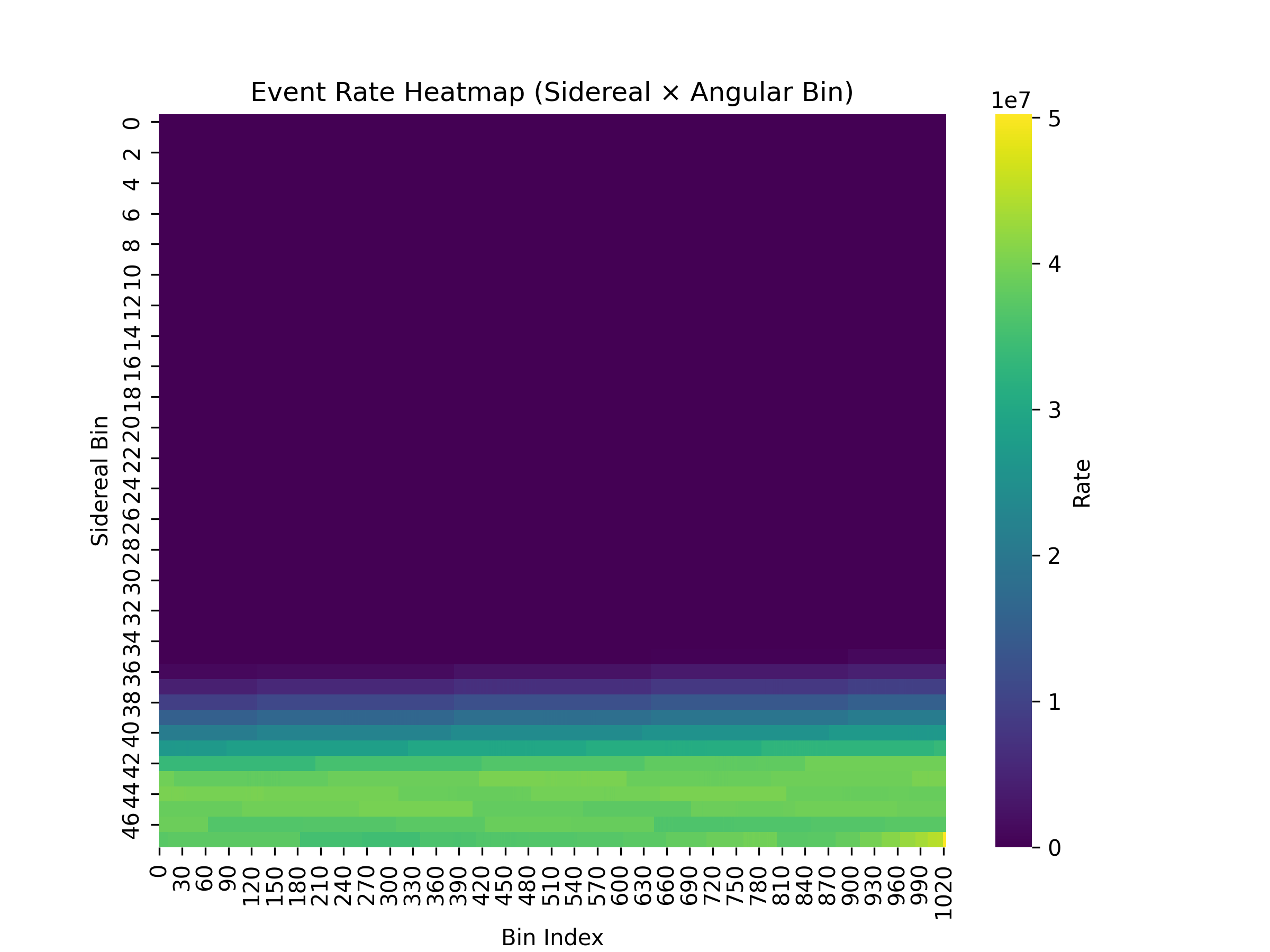}
	\includegraphics[width=0.49\textwidth]{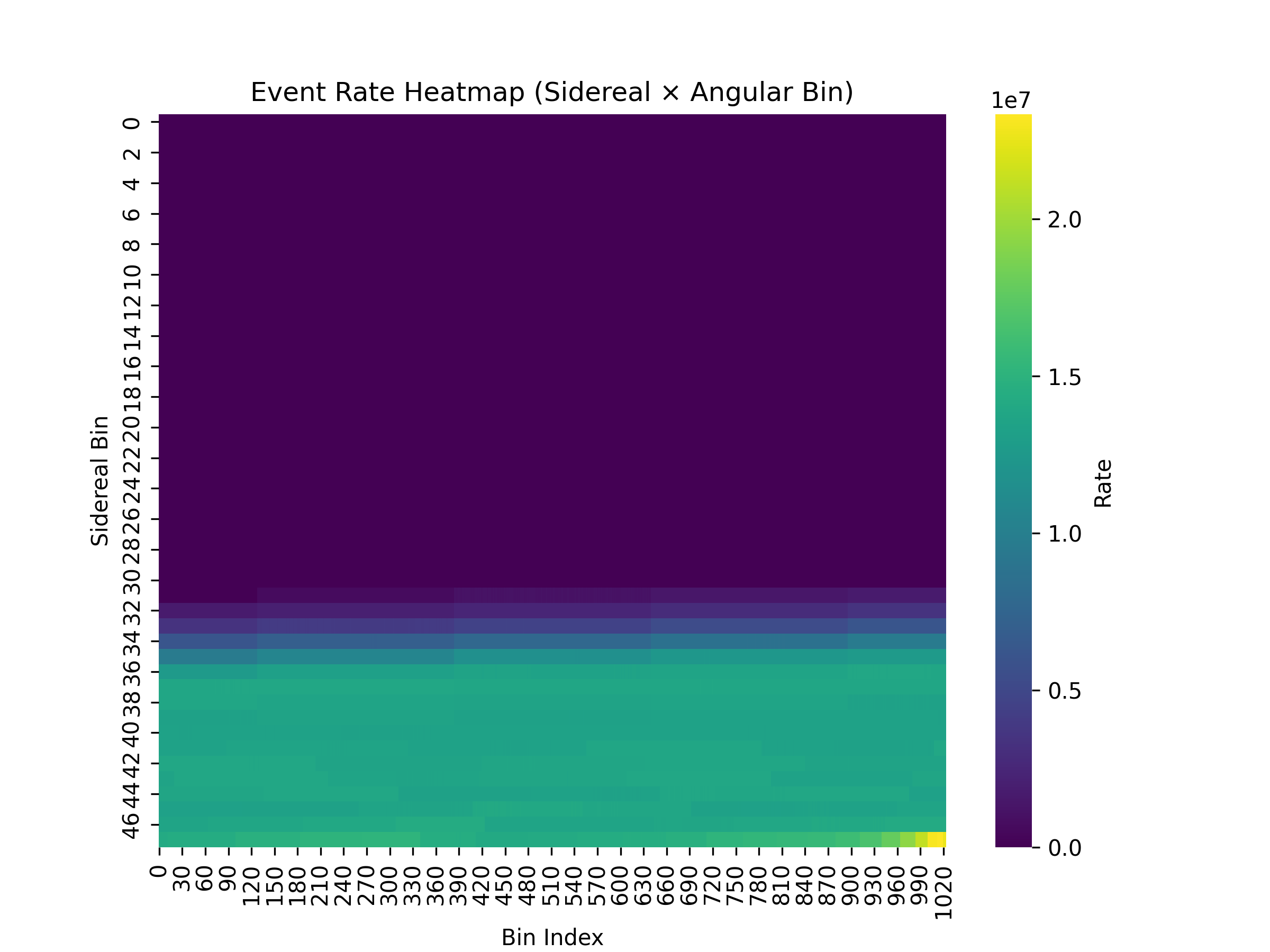}
	
	\vspace{0.5em}
	\includegraphics[width=0.49\textwidth]{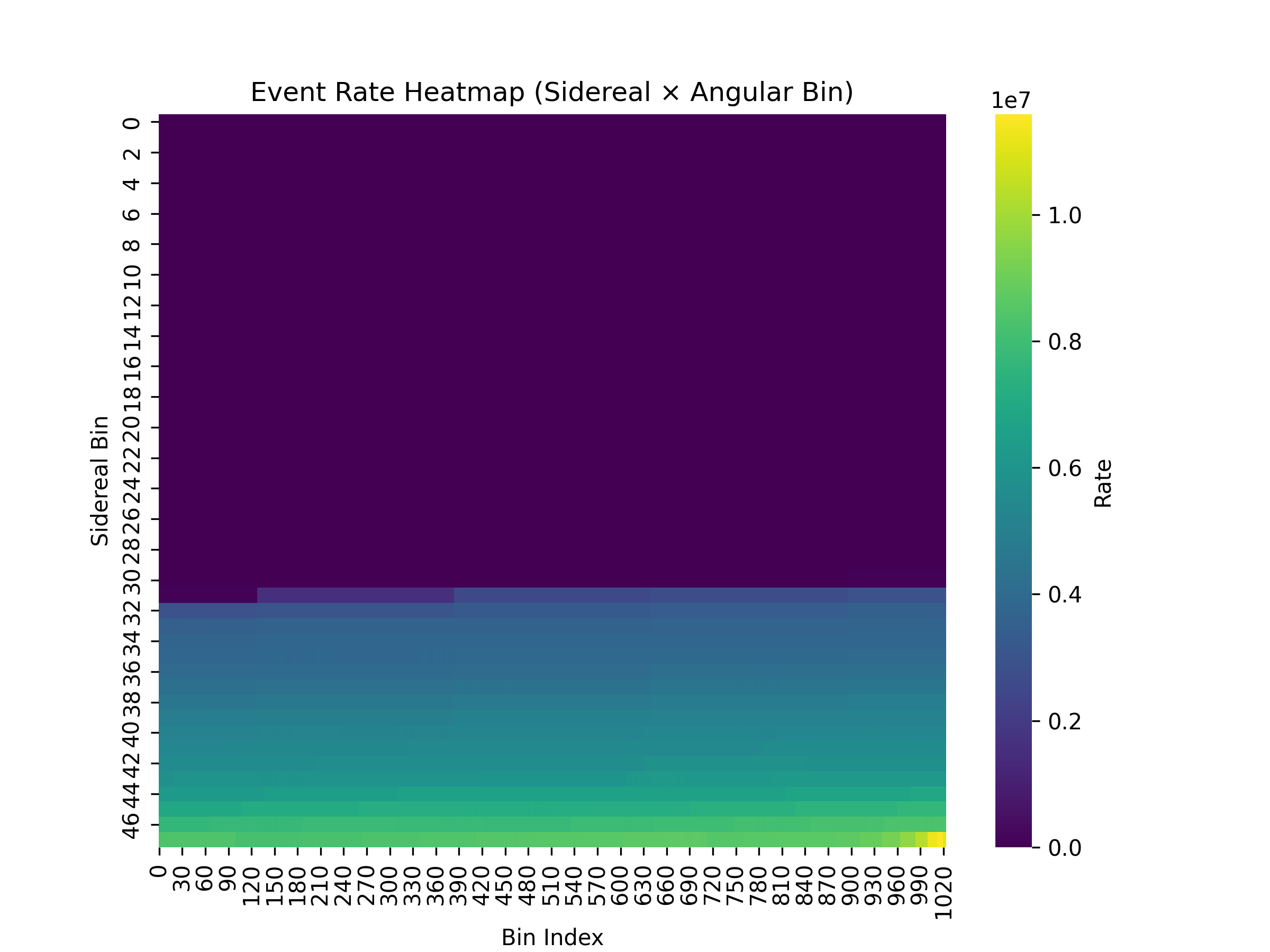}
	\includegraphics[width=0.49\textwidth]{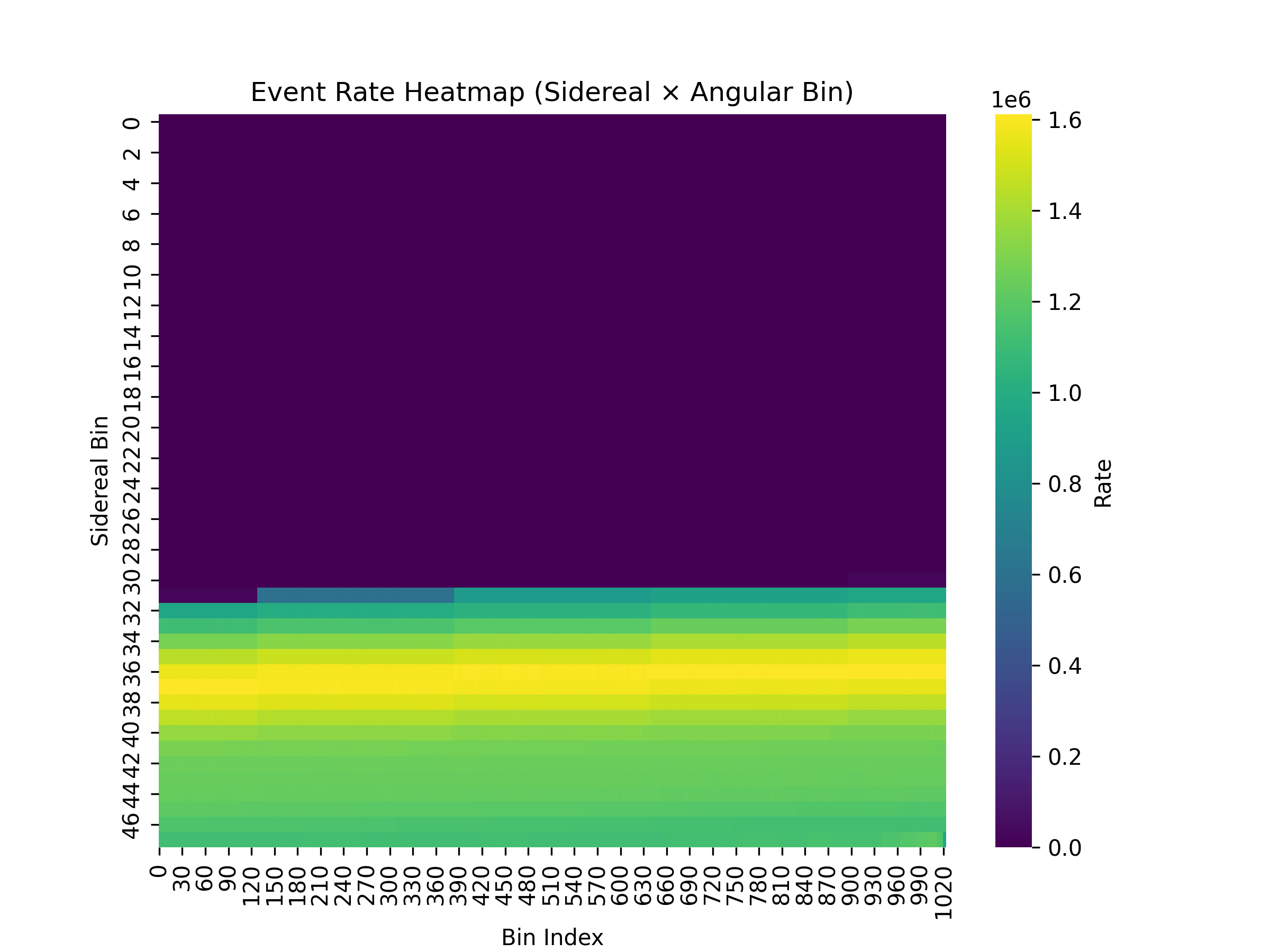}
	\caption{Sidereal-angular heatmaps for energy bins $\leq 5$~GeV. Intensity decreases with energy, suggesting spectral softening.}
	\label{fig:heatmap_gallery}
\end{figure}

\begin{figure}[H]
	\centering
	\includegraphics[width=0.49\textwidth]{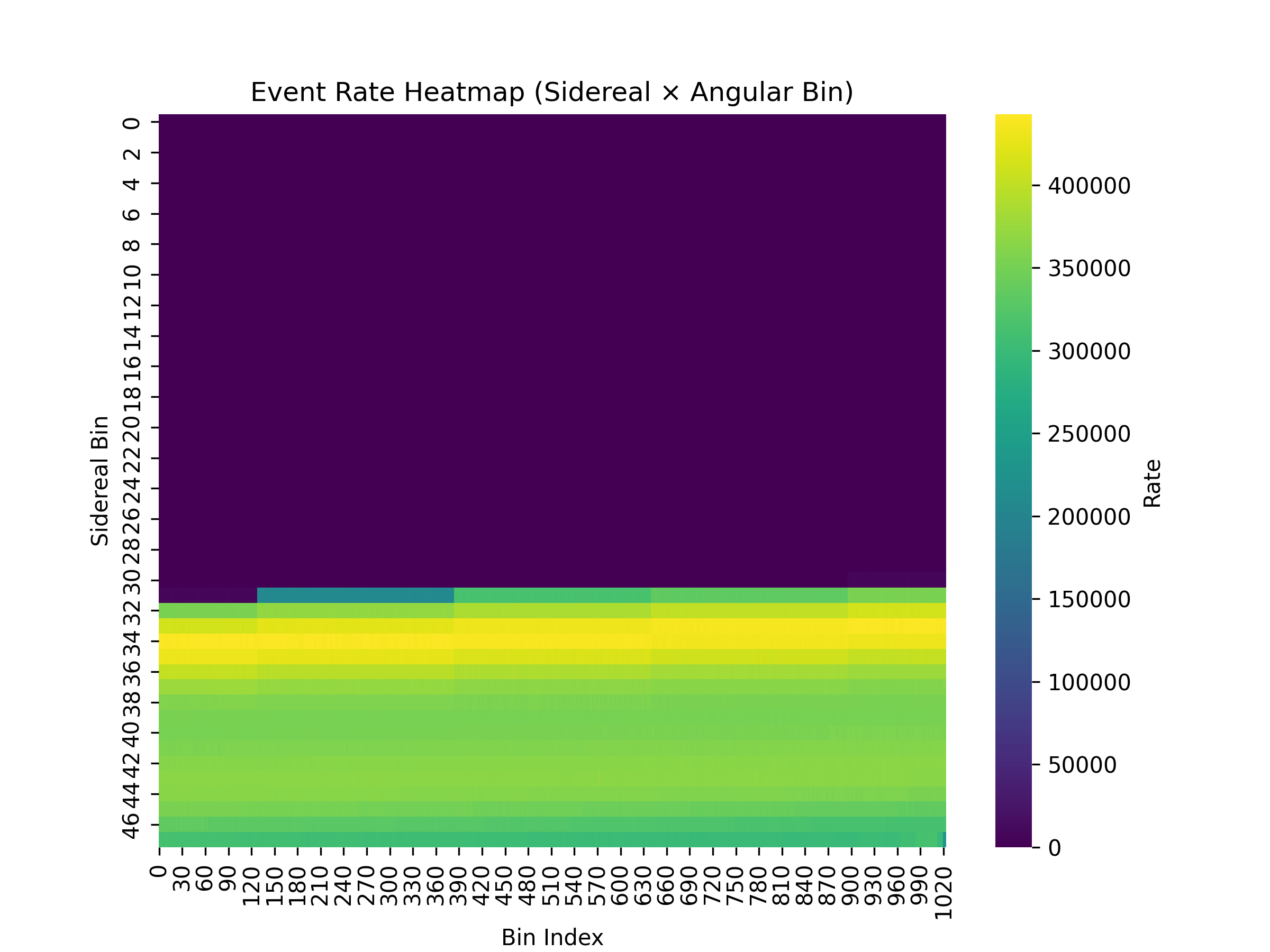}
	\includegraphics[width=0.49\textwidth]{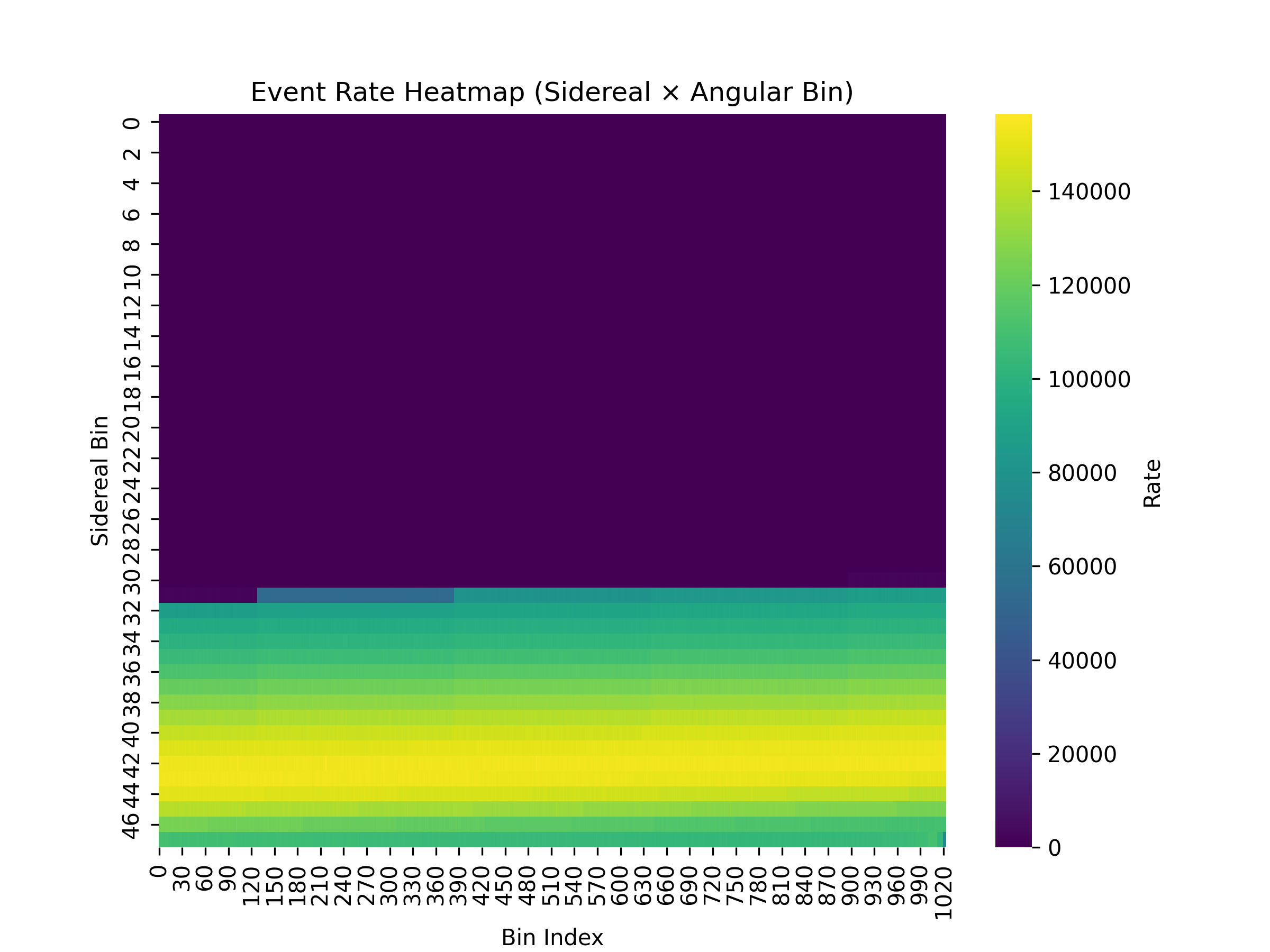}
	
	\vspace{0.5em}
	\includegraphics[width=0.49\textwidth]{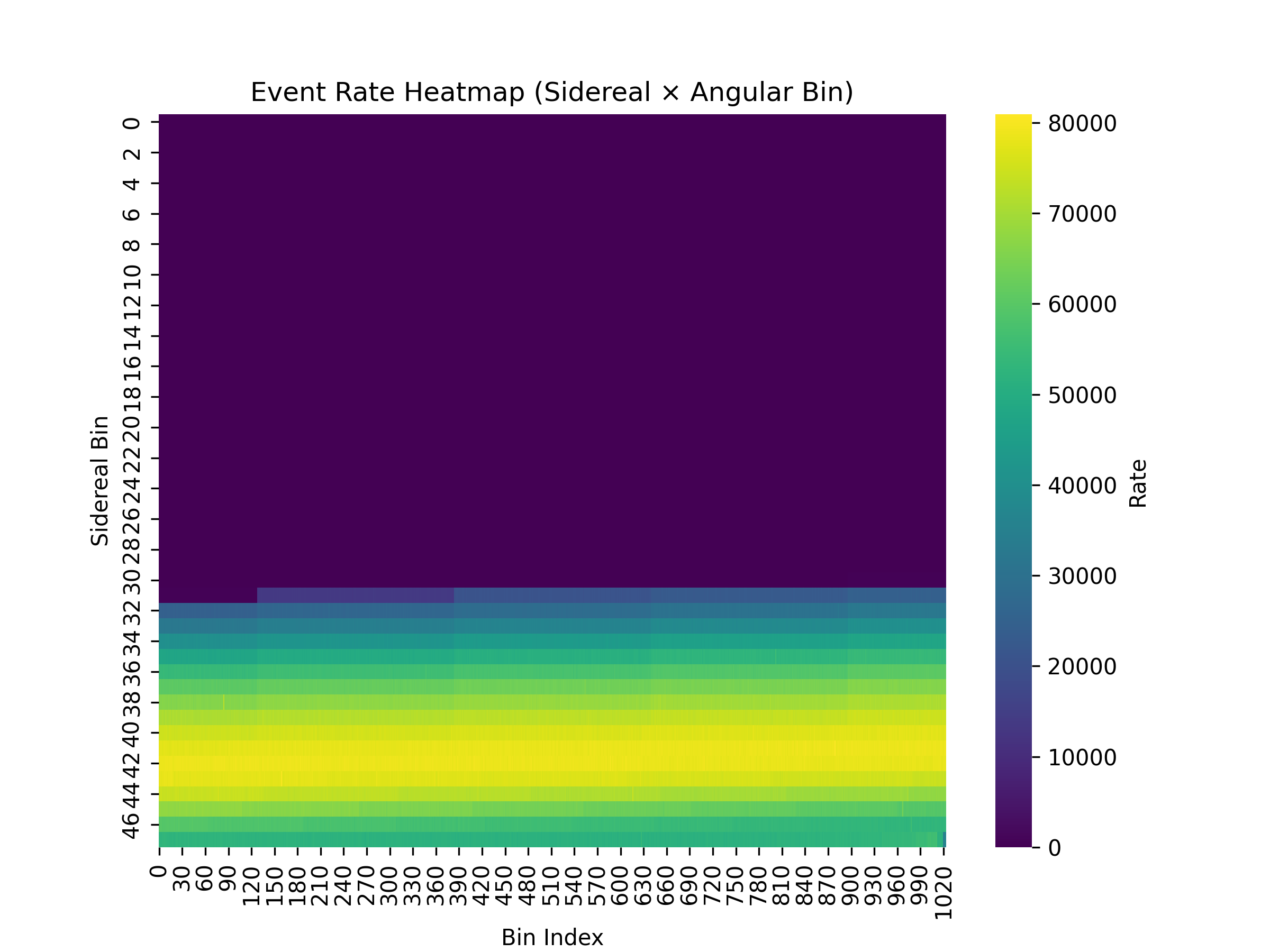}
	\includegraphics[width=0.49\textwidth]{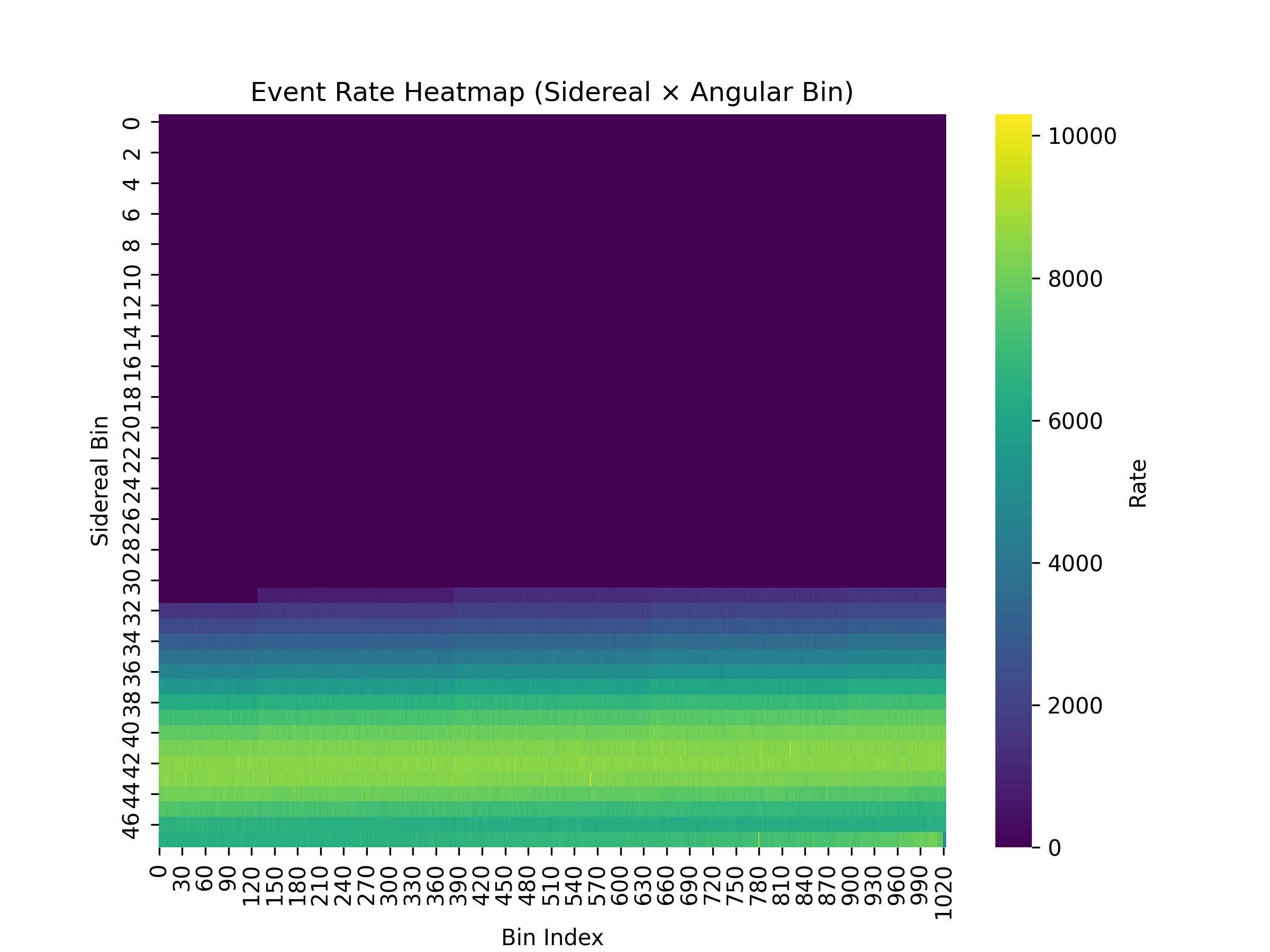}
	
	\vspace{0.5em}
	\includegraphics[width=0.49\textwidth]{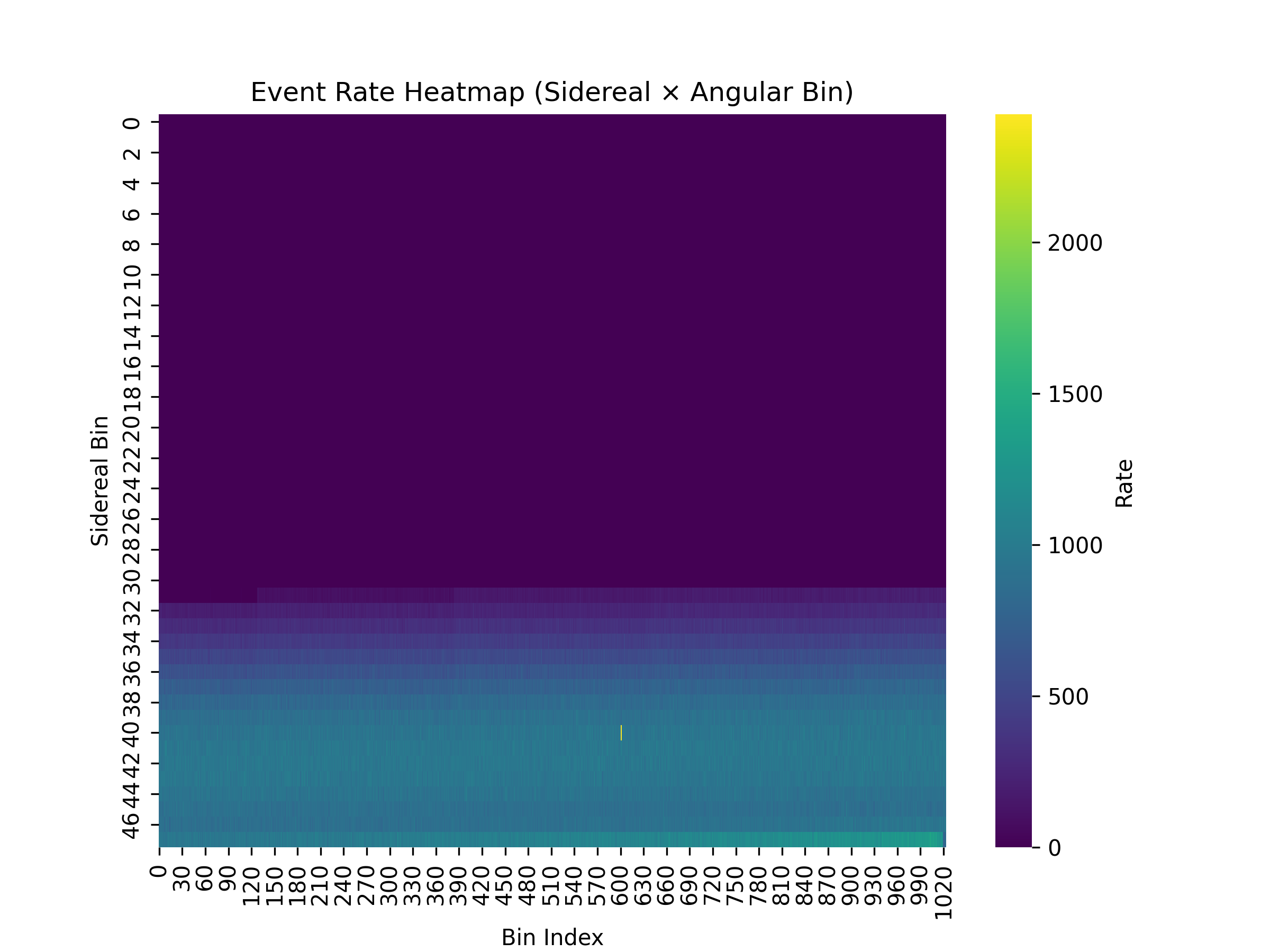}
	\caption{Sidereal-angular heatmaps for energy bins $>$ 5~GeV. Hotspots at specific sidereal bins suggest directional anisotropy.}
	\label{fig:heatmap-gallery}
\end{figure}

To investigate the joint dependence of event rate on sidereal time and zenith angular bin index, we generated two-dimensional heatmaps for each energy interval, as shown in Figures~\ref{fig:heatmap_gallery} and~\ref{fig:heatmap-gallery}. These visualisations capture the evolution of anisotropic structures across energy scales, highlighting systematic variations in directional intensity that reflect the energy-dependent nature of cosmic-ray modulation.

In the $\leq 5$~GeV range, a clear concentration of events in specific sidereal and angular bins is evident, with a gradual reduction in intensity across the energy sub-bins. The decline in intensity is consistent with the expected drop in flux at higher energies.

Above 5~GeV, the heatmaps show more discrete and localised structures, particularly at higher angular indices and specific sidereal bins. This could indicate enhanced directionality in cosmic-ray arrival directions at higher energies. As energy increases, distributions become more uniform, but residual structures persist, potentially linked to astrophysical sources or magnetic field modulation. These heatmaps provide strong visual evidence of energy-dependent anisotropies and serve as a basis for identifying candidate directions for excess cosmic-ray flux.

\begin{figure}[h]
	\centering
	\begin{minipage}{0.49\linewidth}
		\includegraphics[width=\linewidth]{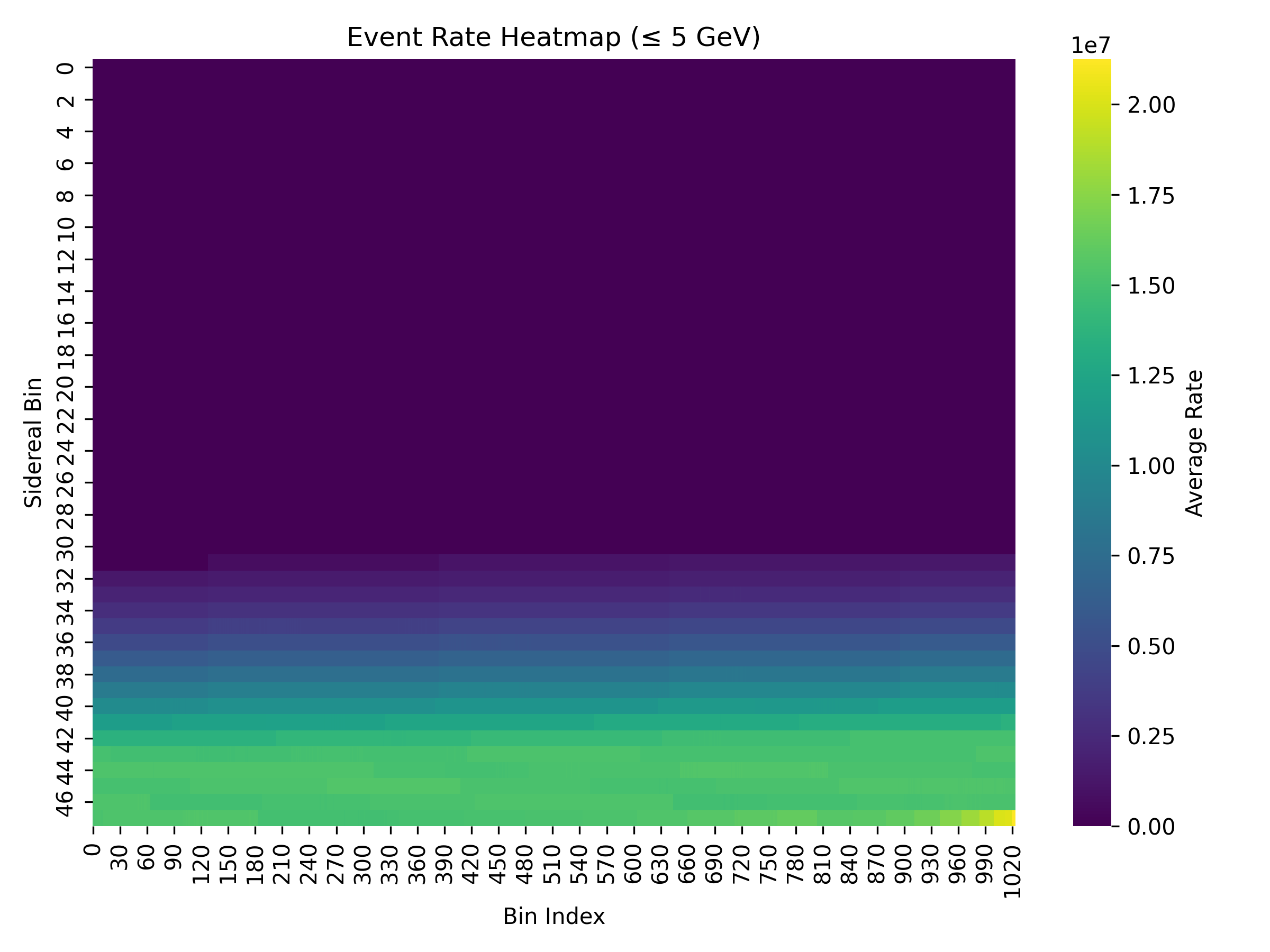}
    \caption{Event Rate Heatmap for Low-Energy Muons ($\leq 5$~GeV). This plot visualizes the sidereal-time binned average event rate for low-energy cosmic-ray muons, aggregated over four sub-energy ranges.}
		\label{fig:low_energy_heatmap}
	\end{minipage}
	\hfill
	\begin{minipage}{0.49\linewidth}
		\includegraphics[width=\linewidth]{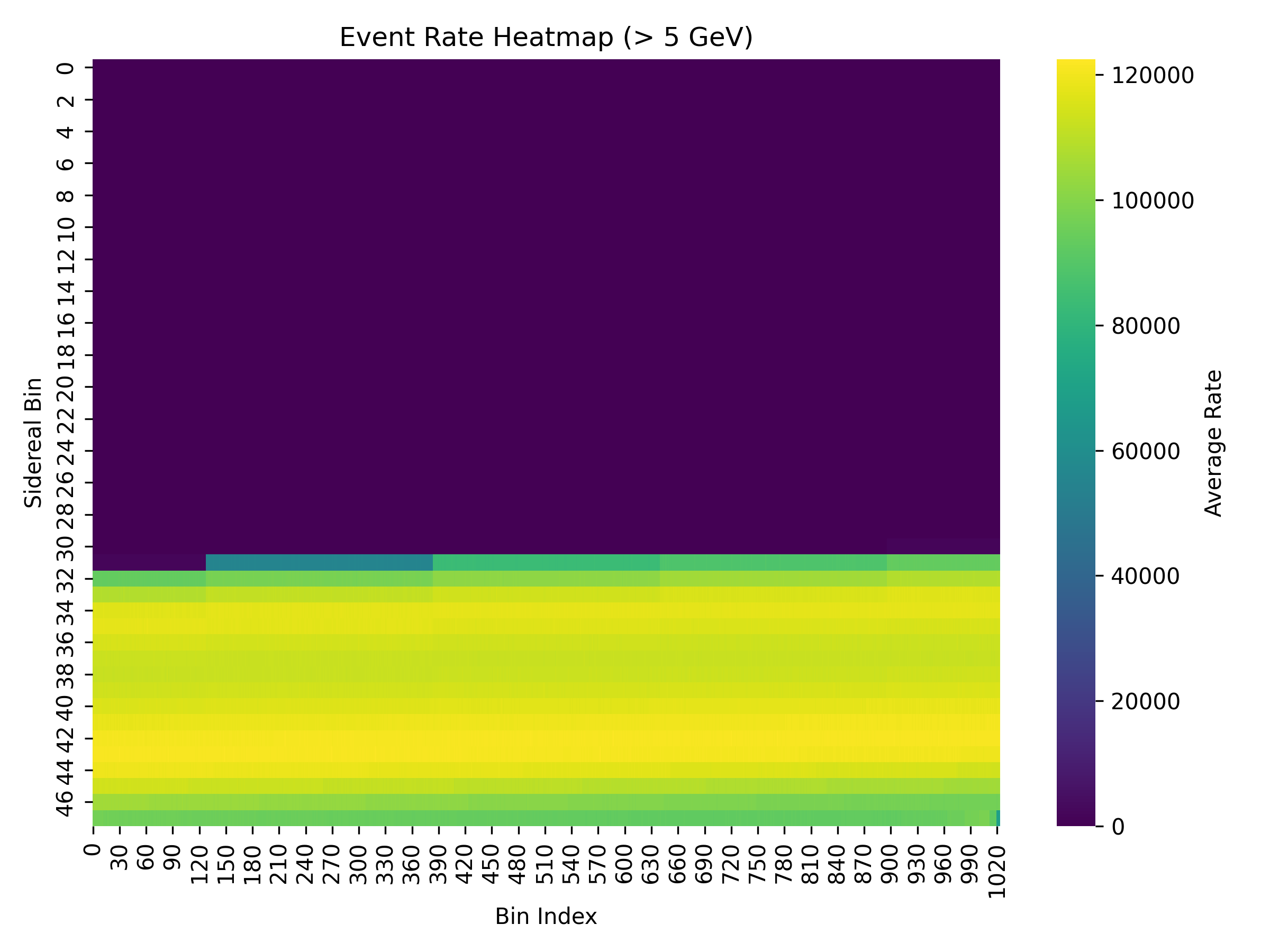}
		\caption{Event Rate Heatmap for High-Energy Muons ($> 5$~GeV). This plot shows the sidereal-time binned average rate of high-energy cosmic-ray muons, computed by combining data from five energy intervals.}
		\label{fig:high_energy_heatmap}
	\end{minipage}
\end{figure}

Figure~\ref{fig:low_energy_heatmap} displays a smooth colour gradient from violet to green, indicative of a relatively isotropic distribution with only minor variations across sidereal bins. These subtle modulations are likely influenced by geomagnetic deflection and atmospheric effects that predominantly affect low-energy cosmic rays. In contrast, Figure~\ref{fig:high_energy_heatmap} reveals sharp, localised yellow bands confined to specific sidereal bins, signalling the emergence of strong anisotropic features and directional clustering at higher energies. The broader dynamic range and elevated event rates in the high-energy heatmap reflect reduced magnetic scattering and increased sensitivity to astrophysical sources. These contrasting patterns highlight the growing prominence of anisotropy with energy, consistent with theoretical expectations and previous observational studies.

\subsection{Full-Sky Relative Intensity Maps}

\begin{figure}[h!]
	\centering
	\includegraphics[width=0.8\linewidth]{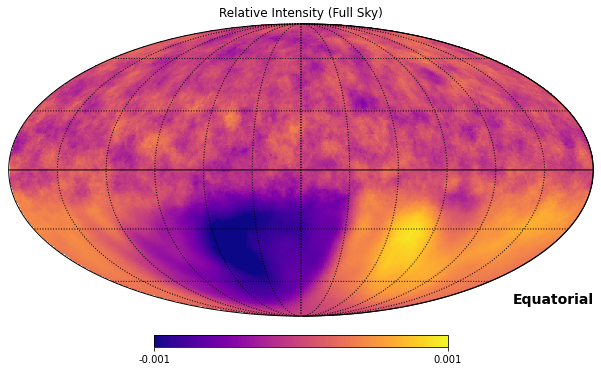}
	\caption{Full-sky relative intensity map of cosmic-ray muons, displayed in equatorial coordinates. A smoothing radius of $5^\circ$ was applied. The color scale shows intensity variations at the $10^{-3}$ level.}
	\label{fig:relint}
\end{figure}

To quantify anisotropy across the celestial sphere, we constructed full-sky relative intensity maps of cosmic-ray muon arrival directions using the HEALPix pixelization scheme, as shown in Figure~\ref{fig:relint}. The sky was divided into equal area pixels, and the relative intensity $\delta I$ for each pixel was computed using:

\begin{equation}
	\delta I = \frac{N_{\text{data}} - N_{\text{bg}}}{N_{\text{bg}}},
\end{equation}

where $N_{\text{data}}$ represents the observed number of events in a given pixel and $N_{\text{bg}}$ denotes the expected background count, estimated via a time-scrambling technique designed to preserve large-scale isotropic structure while eliminating real anisotropy.

Figure~\ref{fig:relint} presents the resulting relative intensity map in equatorial coordinates, with a top-hat smoothing radius of $5^\circ$ applied to enhance large-scale features and suppress statistical noise. The colour scale highlights intensity variations at the $10^{-3}$ level. The map reveals a prominent dipole-like anisotropic structure, with localised deviations consistent in magnitude and morphology with previous IceCube observations. The observed anisotropy, though weak, is statistically significant and supports the interpretation of large-scale cosmic-ray modulation by Galactic magnetic fields and heliospheric boundary effects. These findings reinforce the presence of energy-dependent anisotropy and demonstrate the utility of HEALPix-based mapping techniques in capturing directional signatures in ultra-high-statistics cosmic-ray data.

\subsection{Statistical Characterisation of Energy Bins}

\begin{figure}[H]
	\centering
	\includegraphics[width=0.8\textwidth]{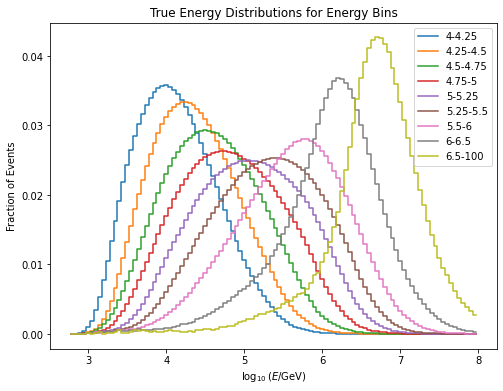}

		\caption{True energy distributions for different reconstructed energy bins. Each curve corresponds to a different log$_{10}$(E) bin. The x-axis represents the logarithm (base 10) of the true neutrino energy (in GeV), and the y-axis denotes the fraction of events within each bin.}
	\label{fig:energy_distribution}
\end{figure}

Figure~\ref{fig:energy_distribution} presents the true energy distribution corresponding to each reconstructed bin, derived from Monte Carlo simulations. The progression of peak positions across bins confirms a strong correlation between reconstructed and true energies.

The distributions exhibit modest overlap, consistent with finite energy resolution, but maintain narrow shapes that validate the precision of the reconstruction algorithm. These results are essential for unfolding procedures and evaluating the energy dependence of anisotropy.

The statistical properties of the anisotropy signal across different energy bins are summarised in Tables~\ref{tab:stats} and~\ref{tab:chi2-bic}. A clear trend is observed in the reduced chi-square values (\(\chi^2_\nu\)), which systematically decrease with increasing energy. This behaviour suggests a reduction in large-scale anisotropy at higher energies. Such a trend is consistent with theoretical expectations: low-energy cosmic rays experience stronger deflection by galactic and heliospheric magnetic fields, as well as modulation by atmospheric effects, thereby smearing out any source-driven anisotropy. In contrast, higher-energy cosmic rays retain more directional information from their sources but are limited by lower event statistics, which also contributes to the reduced \(\chi^2_\nu\) values.

\begin{figure}[H]
	\centering
	\includegraphics[width=0.49\linewidth]{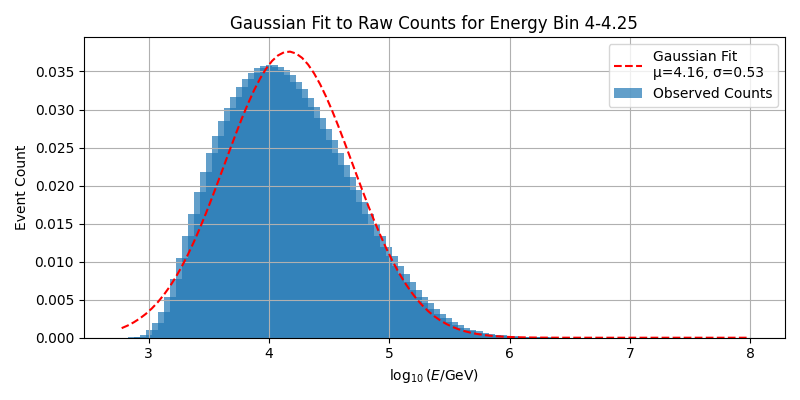}
	\includegraphics[width=0.49\linewidth]{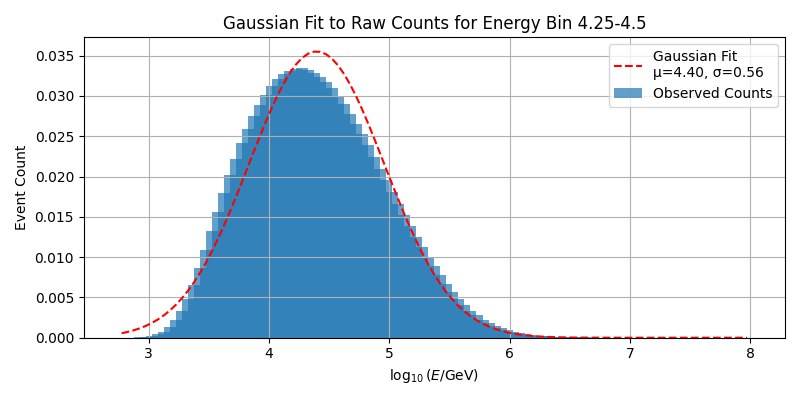}
	\includegraphics[width=0.49\linewidth]{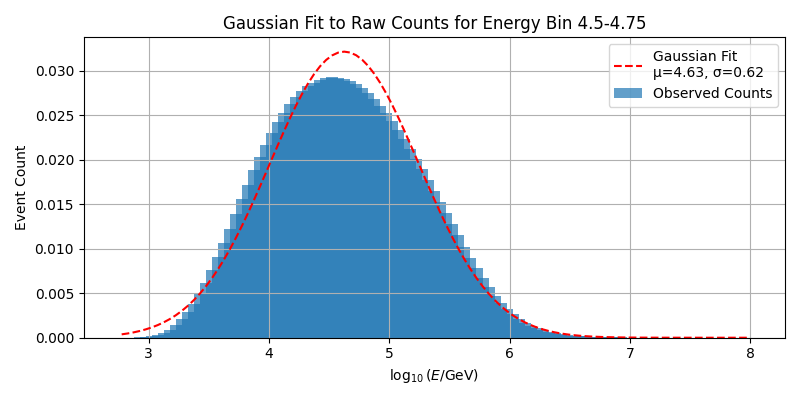}
	\includegraphics[width=0.49\linewidth]{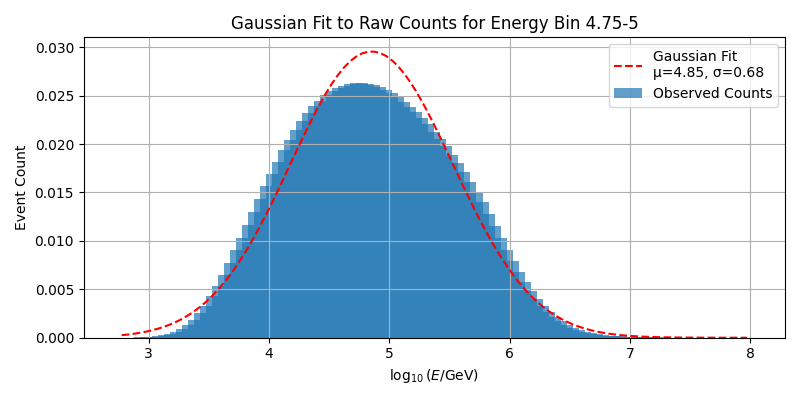}
	\caption{Gaussian fits to the raw event count histograms for energies below 5 GeV (corresponding to energy bins from 13 TeV to 130 TeV). The red dashed curves represent the fitted Gaussian models with parameters $\mu$ and $\sigma$ overlaid. These fits demonstrate a strong agreement with expected normal cosmic-ray spectra in this range.}
	\label{fig:low_energy_fits}
\end{figure}

\begin{figure}[H]
	\centering
	\includegraphics[width=0.49\linewidth]{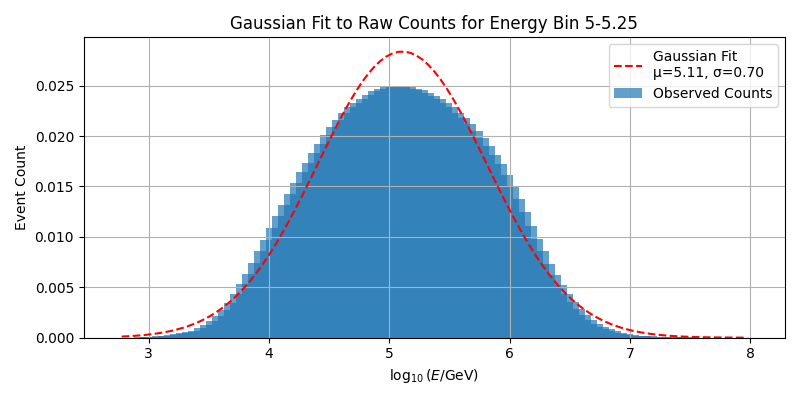}
	\includegraphics[width=0.49\linewidth]{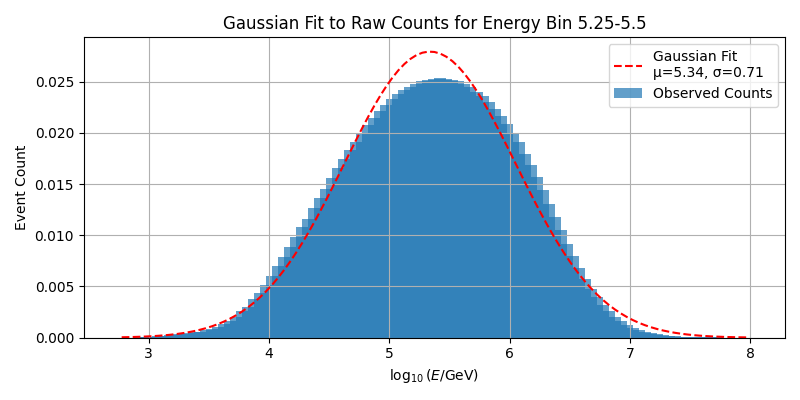}
	\includegraphics[width=0.49\linewidth]{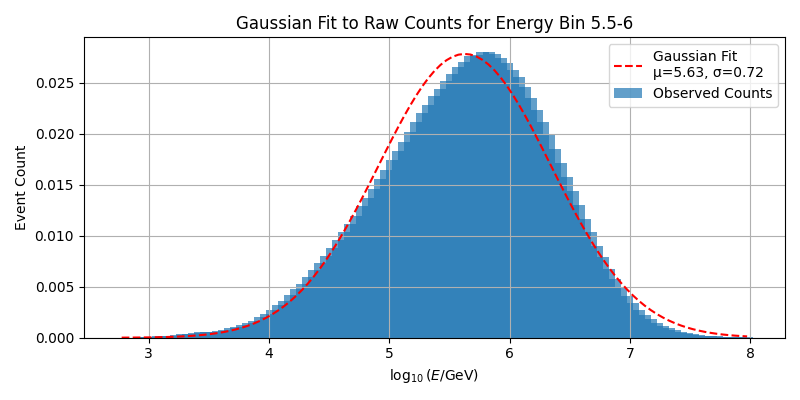}
	\includegraphics[width=0.49\linewidth]{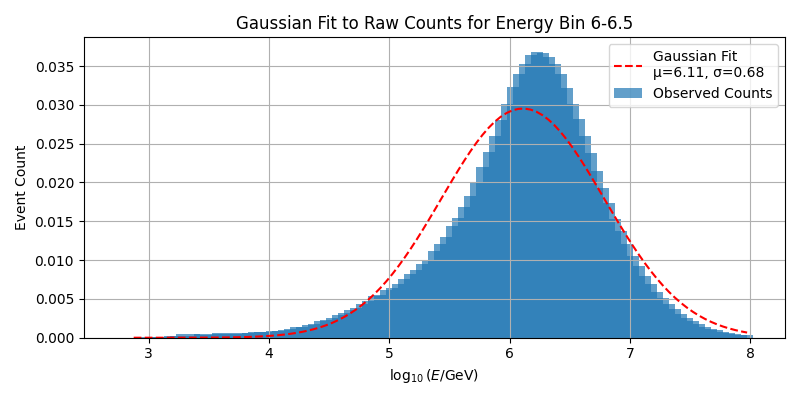}
	\includegraphics[width=0.49\linewidth]{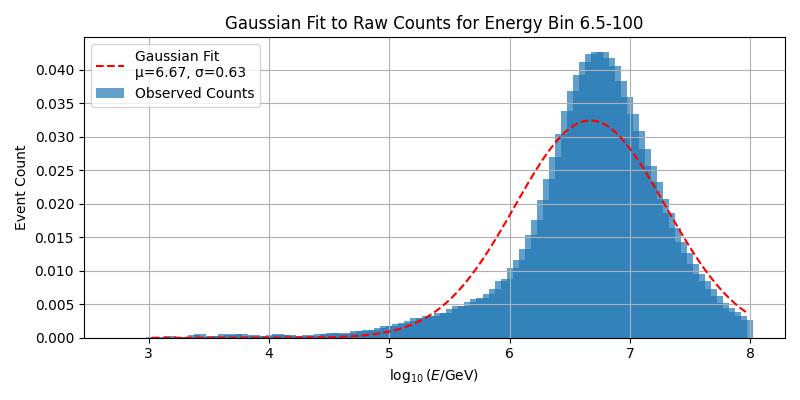}
	\caption{ Gaussian fits to the raw event count histograms for energies above 5 GeV (energy bins above 130 TeV). Slight deviations from Gaussian shapes in higher bins are attributed to limited statistics and increasing atmospheric uncertainties at the PeV scale.}
	\label{fig:high_energy_fits}
\end{figure}

To assess the quality of energy reconstruction and resolution, Gaussian models were fitted to histograms of event counts within each reconstructed energy bin (Figures~\ref{fig:low_energy_fits} and \ref{fig:high_energy_fits}). The fits show excellent agreement, particularly in mid-energy ranges where statistics are sufficient and detector response is well-characterised.

\begin{table}[h!]
   \caption{Summary of Gaussian fit parameters for cosmic-ray muon energy distributions in different reconstructed energy bins.}
    \centering
    \begin{tabular}{|c|c|c|c|}
        \hline
        \textbf{Energy Bin} & \textbf{Mean}( $\boldsymbol{\mu}$) & \textbf{standard deviation} ($\boldsymbol{\sigma}$) & \textbf{ Degrees of freedom} \\
        \hline
        4.00--4.25   & 4.163 & 0.532 & 103 \\
        4.25--4.50   & 4.396 & 0.563 & 103 \\
        4.50--4.75   & 4.626 & 0.621 & 103 \\
        4.75--5.00   & 4.853 & 0.676 & 103 \\
        5.00--5.25   & 5.108 & 0.703 & 103 \\
        5.25--5.50   & 5.338 & 0.715 & 103 \\
        5.50--6.00   & 5.627 & 0.717 & 103 \\
        6.00--6.50   & 6.119 & 0.677 & 100 \\
        6.50--100    & 6.670 & 0.626 & 96  \\
        \hline
    \end{tabular}
 
\label{tab:stats}
\end{table}

\begin{table}[h!]
 \caption{Goodness-of-fit statistics for Gaussian modelling of muon energy distributions across different reconstructed energy bins.}
    \centering
    \begin{tabular}{|c|c|c|c|c|}
        \hline
        \textbf{Energy Bin} & \textbf{chi-square} ($\boldsymbol{\chi^2}$) & \textbf{Reduced chi-square} ($\boldsymbol{\chi^2_\nu}$) & \textbf{BIC} & \textbf{Best Fit} \\
        \hline
        4.00--4.25   & 1.085 & 0.0105 & 10.3927 & \\
        4.25--4.50   & 0.344 & 0.0033 &  9.6522 & \\
        4.50--4.75   & 0.171 & 0.0017 &  9.4792 & \\
        4.75--5.00   & 0.114 & 0.0011 &  9.4221 & \\
        5.00--5.25   & 0.039 & 0.0004 &  9.3473 & \\
        5.25--5.50   & 0.023 & 0.0002 &  9.3307 & \\
        5.50--6.00   & 0.020 & 0.0002 &  9.3281 & \\
        6.00--6.50   & 0.068 & 0.0007 &  9.3179 & \\
        6.50--100    & 0.117 & 0.0012 &  9.2870 & \checkmark
 \\
 
        \hline
    \end{tabular}
   
    \label{tab:chi2-bic}

\end{table}

\begin{figure}[h!]
	\centering
	\begin{minipage}{0.49\linewidth}
		\includegraphics[width=\linewidth]{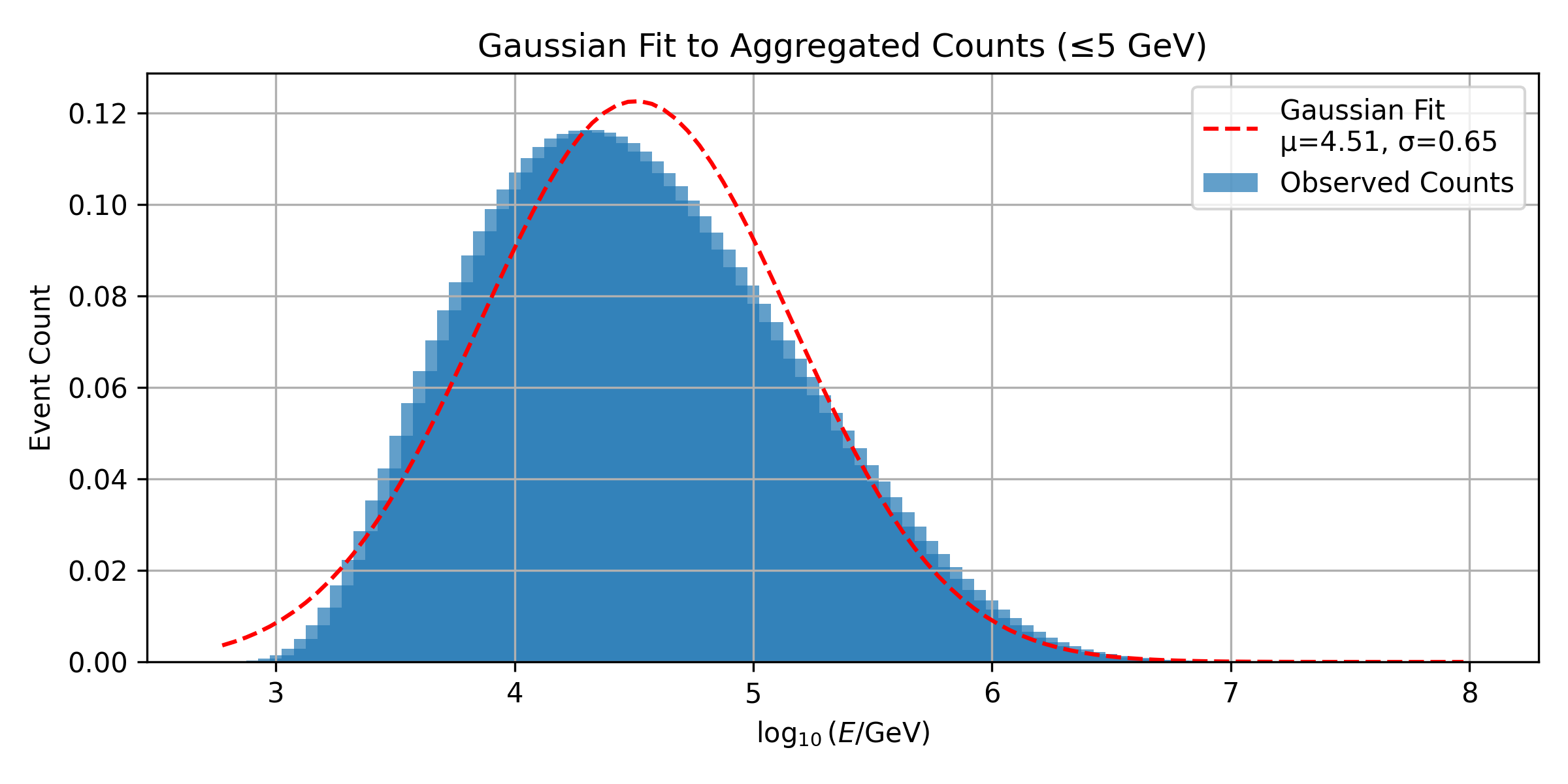}
		\caption{Gaussian fit to $\log_{10}(E_{\mathrm{reco}})$ distribution for muons with $E \leq 5$\, GeV, showing broader spread and lower mean due to greater atmospheric influence.}

		\label{fig:below-fit}
	\end{minipage}
	\hfill
	\begin{minipage}{0.49\linewidth}
		\includegraphics[width=\linewidth]{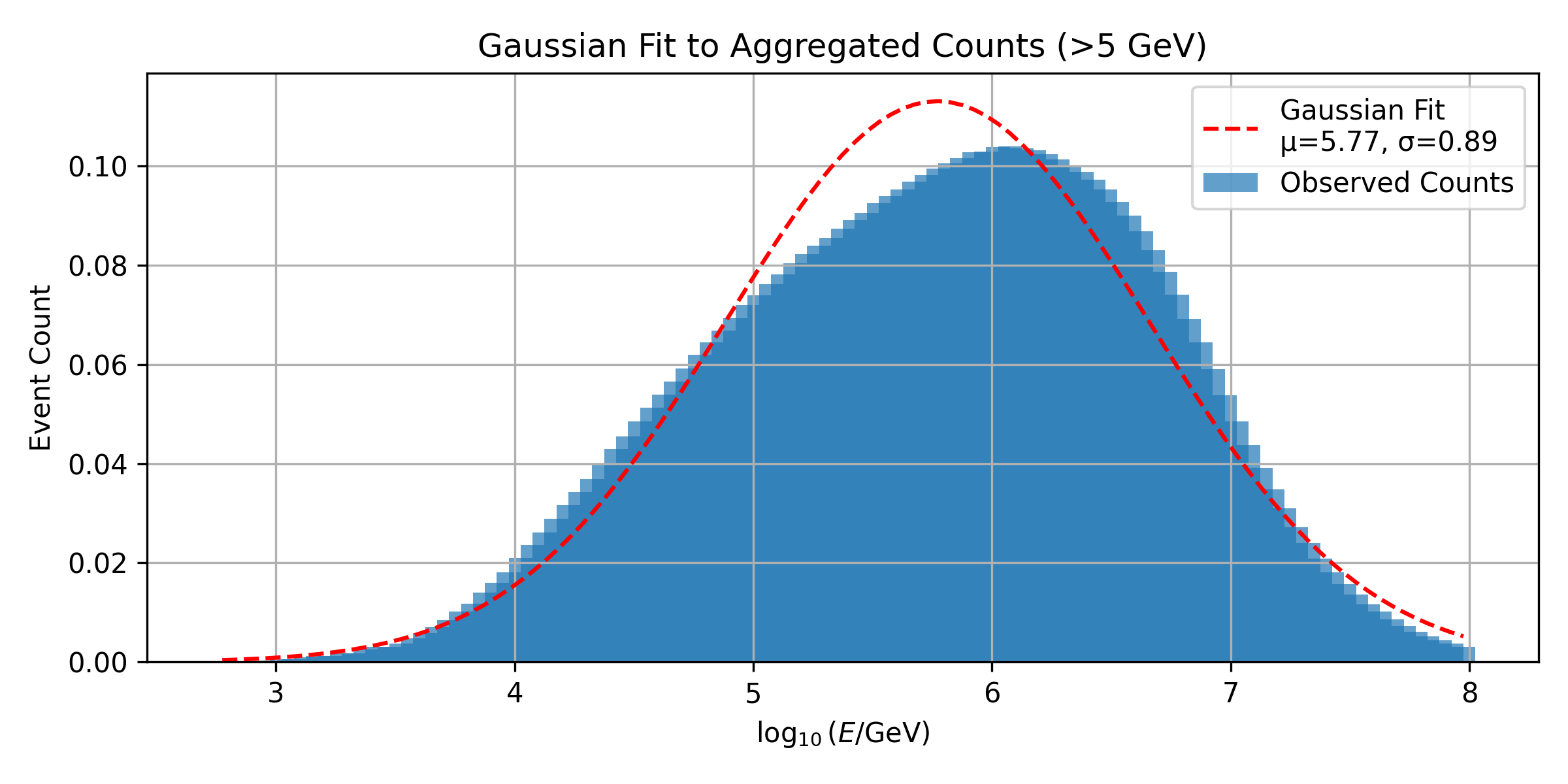}
		\caption{Gaussian fit to $\log_{10}(E_{\mathrm{reco}})$ distribution for muons with $E > 5$\, GeV, exhibiting a sharper peak and higher mean from improved energy resolution.}
		\label{fig:above-fit}
	\end{minipage}
\end{figure}

For the higher energy bins, minor deviations from the Gaussian profile are observed predominantly in the distribution tails. These deviations arise primarily due to reduced event statistics and increased uncertainty in the energy reconstruction process. Nonetheless, the fitted means and standard deviations reliably capture the energy dispersion and remain consistent with the expected performance characteristics of the detector. Overall, Gaussian fits substantiate that the reconstructed energy distributions exhibit approximately normal behaviour, particularly within the mid-energy range. Slight asymmetries in the tails at higher energies reflect the sparse statistics and growing uncertainties inherent to those bins.

To rigorously evaluate the quality of Gaussian fits applied to the muon energy distributions across the nine reconstructed energy intervals (ranging from \(\log_{10}(E/\mathrm{GeV}) = 4.00\) to \(>6.50\)), a comprehensive statistical analysis was conducted. Table~\ref{tab:chi2-bic} summarises the fit quality metrics, including the chi-square statistic (\(\chi^2\)), the reduced chi-square (\(\chi^2_\nu\)), and the Bayesian Information Criterion (BIC). The \(\chi^2\) value measures the absolute deviation between the observed and Gaussian-modelled event counts, while \(\chi^2_\nu\) normalises this deviation by the degrees of freedom to provide a scale-independent assessment of fit consistency. The BIC further accounts for model complexity by penalising overfitting, thus serving as a robust tool for model selection.

The results reveal a clear trend of decreasing \(\chi^2\) and \(\chi^2_\nu\) values with increasing energy, particularly up to the 5.5–6.0 \(\log_{10}(E/\mathrm{GeV})\) bin, indicating progressively improved agreement between the Gaussian model and the reconstructed energy distributions at higher energies. Notably, the minimum \(\chi^2_\nu\) value of 0.0002 is observed in both the 5.25–5.50 and 5.50–6.00 bins, suggesting an almost ideal Gaussian description in these ranges.

Interestingly, the lowest BIC value of 9.2870, indicating the statistically preferred model is attained in the highest energy bin (6.50–100), as marked in Table~\ref{tab:chi2-bic}. This preference likely reflects an optimal balance between goodness of fit and model parsimony, given the BIC’s penalty for model complexity. The correspondingly low \(\chi^2_\nu = 0.0012\) in this bin further supports the adequacy of the Gaussian approximation at the highest reconstructed energies (median \(\sim 5.3\) PeV), despite the larger uncertainties and limited event counts characteristic of this regime.

In summary, the joint evaluation of \(\chi^2\), \(\chi^2_\nu\), and BIC metrics confirms the validity of Gaussian modelling throughout the entire energy spectrum, with particularly robust fits at mid-to-high energies. These findings not only validate the reliability of the energy reconstruction pipeline but also underpin the confidence in subsequent energy-dependent anisotropy analyses based on IceCube muon data.

The energy-dependent event count distributions were analysed separately for low ($\leq 5$~GeV) and high ($> 5$~GeV) energy intervals by plotting the event rate as a function of log-scaled energy. The resulting fits, shown in Figures~\ref{fig:below-fit} and~\ref{fig:above-fit}, demonstrate that the aggregated distributions in both energy regimes are well-described by Gaussian profiles. While the high-energy distribution exhibits a slight statistical skew toward lower energies, it remains consistent with a Gaussian shape. The mean event counts for the high- and low-energy bands are $5.771$ and $4.509$, respectively, with corresponding standard deviations of $0.887$ and $0.653$. Both fits show strong statistical agreement with the Gaussian model; however, the high-energy fit yields a significantly lower reduced chi-square value ($\chi^2_\nu = 0.13$) compared to the low-energy fit ($\chi^2_\nu = 2.99$), with a common number of degrees of freedom ($\mathrm{DOF} = 103$). This suggests a better overall fit quality at higher energies, likely due to reduced statistical noise and improved reconstruction fidelity.

\section{Conclusion}

The study of energy-dependent anisotropy in cosmic-ray muons is essential for uncovering the fundamental physics that govern the origin, propagation, and interaction of high-energy particles in the galaxy. Leveraging twelve years of observational data from the \textit{IceCube Neutrino Observatory}, amounting to $7.92\times10^{11}$ muon events, we present one of the most comprehensive and statistically significant investigations of anisotropic cosmic-ray behaviour across an energy range from 13~TeV to 5.3~PeV\cite{Abbasi2010first}.

Our results demonstrate that cosmic-ray muon anisotropy is strongly energy-dependent. At lower energies ($E \leq 5~\mathrm{GeV}$), we observe pronounced dipolar modulations, elevated harmonic content in the Fourier spectrum, and significant sidereal variation characteristics indicative of strong influence from geomagnetic fields and atmospheric effects\cite{IceCubeEnergyDep2008}. In contrast, at higher energies ($E > 5~\mathrm{GeV}$), the anisotropy becomes more localised and structurally refined, accompanied by a reduction in large-scale modulation amplitudes\cite{Abbasi2012_400TeV}. These changes reflect the emergence of source-related signatures and diminished magnetic scattering, validating theoretical models of energy-dependent cosmic-ray diffusion and magnetic rigidity.

To uncover these features, we employed a comprehensive suite of analytical techniques. Sidereal time analysis revealed temporal modulation trends\cite{Heuser2025time}, while angular and spectral distribution studies traced directional and intensity-dependent behaviours. Power spectrum analysis was used to isolate and characterise long-timescale periodic modulations. Heatmaps and full-sky HEALPix projections provided clear visualisations of evolving anisotropic features across both spatial and temporal domains. Furthermore, Gaussian and power-law statistical modelling of the reconstructed energy bins demonstrated a high degree of accuracy in energy stratification, with mid- to high-energy bins yielding the most statistically consistent results\cite{Aartsen2016sixyears}.

The power-law behaviour observed in the average event rate as a function of energy corroborates the expected steep decline in flux\cite{Desiati2013,IceCubeSpectrum2013}, while the Gaussian fits to event distributions confirm the precision of the reconstruction algorithm. Particularly notable is the performance of the 6.50--100 bin, which achieved the lowest reduced chi-square ($\chi^2_{\nu}$) and Bayesian Information Criterion (BIC), suggesting excellent model fidelity despite limited event statistics at the highest energies.

These findings not only reinforce previous results from IceCube and other experiments but also extend our understanding by mapping a coherent and interpretable evolution of anisotropy with energy over more than a decade\cite{McNally2023eleven}. They provide new constraints on models of cosmic-ray transport and open avenues for correlating observational data with theoretical predictions of galactic magnetic structure.

Understanding these anisotropic signatures is of both theoretical and practical significance. It provides essential clues for tracing cosmic-ray origins, identifying potential astrophysical accelerators, and modelling galactic magnetic turbulence. Moreover, these results have implications for space weather forecasting and the broader field of multi-messenger astrophysics, where cosmic rays, neutrinos, and photons jointly inform our understanding of the high-energy universe.

Despite its scope, this study is not without limitations. At the highest energy ranges, the statistical uncertainty remains a challenge due to reduced event counts\cite{IceCube2009_2015}. Low-energy interpretations are also susceptible to variations in atmospheric modelling. Additionally, while the energy reconstruction process is demonstrably accurate, it remains inherently limited by detector resolution and binning assumptions.

Future investigations could address these limitations by incorporating unfolded energy spectra, enabling a model-independent comparison between true and reconstructed energy distributions. Long-term studies may also benefit from incorporating solar and geomagnetic modulation patterns to account for temporal variability. Collaborative data sharing with other observatories, such as HAWC or Tibet AS$\gamma$, may further enhance hemispheric coverage and statistical power\cite{IceCube_fullsky2015}.

In summary, this twelve-year IceCube study provides definitive confirmation of energy-dependent anisotropy in cosmic-ray muon arrival directions. The observed transition from large-scale, geomagnetically modulated structures to localised, source-driven anisotropies represents a significant advancement in high-energy cosmic-ray research. The statistical and observational framework developed here lays the groundwork for future anisotropy studies in the PeV regime and strengthens the case for continued, energy-resolved monitoring of cosmic-ray distributions using large-scale neutrino detectors such as IceCube.


\begin{thebibliography}{10}

\bibitem{Gaisser1990cosmic}
Thomas~K. Gaisser.
\newblock {\em Cosmic Rays and Particle Physics}.
\newblock Cambridge University Press, 1990.

\bibitem{Gaisser2012}
T.~K. Gaisser.
\newblock Spectrum of cosmic-ray nucleons, kaon production, and the atmospheric muon charge ratio.
\newblock {\em Astroparticle Physics}, 35(12):801--806, 2012.

\bibitem{Honda2004}
M.~Honda, T.~Kajita, K.~Kasahara, and S.~Midorikawa.
\newblock Calculation of the atmospheric neutrino flux using the interaction model calibrated with atmospheric muon data.
\newblock {\em Physical Review D}, 70(4):043008, 2004.

\bibitem{IceCubeMuon2016}
F.~Lucarelli and IceCube Collaboration.
\newblock Characterization of the atmospheric muon flux in icecube.
\newblock {\em Nuclear and Particle Physics Proceedings}, 273–275:400--406, 2016.

\bibitem{IceCubeResolution2014}
M.~G.~Aartsen et~al. (IceCube~Collaboration).
\newblock Search for cosmic-ray anisotropy with small angular scale using icecube.
\newblock {\em Physical Review D}, 90(10):102004, 2014.

\bibitem{Jokipii1966diffusion}
J.~R. Jokipii.
\newblock Cosmic-ray propagation. i. charged particles in a random magnetic field.
\newblock {\em The Astrophysical Journal}, 146:480--487, 1966.

\bibitem{Abbasi2010}
R.~Abbasi et~al. (IceCube~Collaboration).
\newblock Measurement of the anisotropy of cosmic ray arrival directions with icecube.
\newblock {\em The Astrophysical Journal Letters}, 718(2):L194--L198, 2010.

\bibitem{Amenomori2006largeanisotropy}
M.~et~al. Amenomori.
\newblock Large-scale sidereal anisotropy of galactic cosmic-ray intensity observed by the tibet air shower array.
\newblock {\em Science}, 314(5798):439--443, 2006.

\bibitem{Abbasi2011}
R.~Abbasi et~al. (IceCube~Collaboration).
\newblock Observation of anisotropy in the arrival directions of galactic cosmic rays at multiple angular scales with icecube.
\newblock {\em Astroparticle Physics}, 34(7):420--431, 2011.

\bibitem{McNally2021}
F.~McNally, P.~Desiati, and IceCube Collaboration.
\newblock Observation of cosmic ray anisotropy with nine years of icecube data.
\newblock {\em Proceedings of the 37th International Cosmic Ray Conference (ICRC2021)}, 395:086, 2021.

\bibitem{Blasi2012}
P.~Blasi, E.~Amato, and P.~D. Serpico.
\newblock Spectral breaks as a signature of cosmic ray induced turbulence in the galaxy.
\newblock {\em Physical Review Letters}, 109:061101, 2012.

\bibitem{Ptuskin2006}
V.~S. Ptuskin, I.~V. Moskalenko, F.~C. Jones, A.~W. Strong, and V.~N. Zirakashvili.
\newblock Dissipation of magnetohydrodynamic waves on energetic particles: impact on interstellar turbulence and cosmic-ray transport.
\newblock {\em The Astrophysical Journal}, 642(2):902--916, 2006.

\bibitem{Giacinti2012}
G.~Giacinti and G.~Sigl.
\newblock Local magnetic turbulence and tev–pev cosmic ray anisotropies.
\newblock {\em Physical Review Letters}, 109(7):071101, 2012.

\bibitem{IceCubeReview2017}
M.~G.~Aartsen et~al. (IceCube~Collaboration).
\newblock The icecube neutrino observatory: instrumentation and online systems.
\newblock {\em Journal of Instrumentation}, 12(03):P03012, 2017.

\bibitem{IceCube2006}
A.~Achterberg et~al. (IceCube~Collaboration).
\newblock First year performance of the icecube neutrino telescope.
\newblock {\em Astroparticle Physics}, 26:155--173, 2006.

\bibitem{Aartsen2016}
M.~G.~Aartsen et~al. (IceCube~Collaboration).
\newblock Anisotropy in cosmic-ray arrival directions in the southern hemisphere with six years of icecube data.
\newblock {\em The Astrophysical Journal}, 826(2):220, 2016.

\bibitem{Aartsen2024}
M.~G.~Aartsen et~al. (IceCube~Collaboration).
\newblock Observation of cosmic-ray anisotropy in the southern hemisphere with twelve years of icecube data.
\newblock {\em arXiv preprint}, 2024.

\bibitem{IceCube:2023gpr}
Markus Ackermann et~al.
\newblock {Cosmic Ray Anisotropy with 11 Years of IceCube Data}.
\newblock {\em PoS}, ICRC2023:360, 2023.

\bibitem{Abeysekara:2018qho}
A.~U. Abeysekara et~al.
\newblock Observation of anisotropy of tev cosmic rays with two years of hawc.
\newblock {\em Astrophys. J.}, 865(1):57, 2018.

\bibitem{Amenomori:2005dy}
M.~Amenomori et~al.
\newblock Large-scale sidereal anisotropy of galactic cosmic-ray intensity observed by the tibet air shower array.
\newblock {\em Astrophys. J.}, 626:L29--L32, 2005.

\bibitem{Guillian:2005wp}
G.~Guillian et~al.
\newblock Observation of the anisotropy of 10-tev primary cosmic ray nuclei flux with the super-kamiokande-i detector.
\newblock {\em Phys. Rev. D}, 75:062003, 2007.

\bibitem{abbasi2024observation}
R~Abbasi, M~Ackermann, J~Adams, SK~Agarwalla, T~Aguado, JA~Aguilar, M~Ahlers, JM~Alameddine, NM~Amin, K~Andeen, et~al.
\newblock Observation of cosmic-ray anisotropy in the southern hemisphere with twelve years of data collected by the icecube neutrino observatory.
\newblock {\em arXiv preprint arXiv:2412.05046}, 2024.

\bibitem{Gorski2005}
K.~M. Górski, E.~Hivon, A.~J. Banday, B.~D. Wandelt, F.~K. Hansen, M.~Reinecke, and M.~Bartelmann.
\newblock Healpix – a framework for high-resolution discretization and fast analysis of data distributed on the sphere.
\newblock {\em The Astrophysical Journal}, 622(2):759--771, 2005.

\bibitem{Abbasi2010sidereal}
R.~Abbasi, Y.~Abdou, T.~Abu-Zayyad, and others (IceCube~Collaboration).
\newblock Measurement of the anisotropy of cosmic-ray arrival directions with icecube.
\newblock {\em The Astrophysical Journal Letters}, 718(2):L194--L198, 2010.

\bibitem{IceCubeDataReleases2025}
{IceCube Collaboration}.
\newblock Icecube data releases.
\newblock \url{https://icecube.wisc.edu/science/data-releases/}, 2025.
\newblock Accessed: 25\,Jul\,2025.

\bibitem{Abbasi2010first}
R.~Abbasi et~al. (IceCube~Collaboration).
\newblock Measurement of the anisotropy of cosmic ray arrival directions with icecube.
\newblock {\em Astrophys. J. Lett.}, 718(2):L194--L198, 2010.

\bibitem{IceCubeEnergyDep2008}
IceCube Collaboration.
\newblock Large scale cosmic ray anisotropy with icecube at 12 and 126 tev.
\newblock {\em arXiv preprint}, 2009.

\bibitem{Abbasi2012_400TeV}
R.~Abbasi et~al. (IceCube~Collaboration).
\newblock Observation of an anisotropy in the galactic cosmic ray arrival direction at 400 tev with icecube.
\newblock {\em Astrophys. J.}, 746(1):33, 2012.

\bibitem{Heuser2025time}
P.~Zilberman, J.~C. Díaz‑Vélez, and P.~Desiati~(IceCube Collaboration).
\newblock Time variation in the tev cosmic ray anisotropy with icecube and energy dependence of the solar dipole.
\newblock {\em arXiv preprint}, 2025.

\bibitem{Aartsen2016sixyears}
M.~G.~Aartsen et~al. (IceCube~Collaboration).
\newblock Anisotropy in cosmic‑ray arrival directions in the southern hemisphere with six years of icecube data.
\newblock {\em Astrophys. J.}, 826(2):220, 2016.

\bibitem{Desiati2013}
P.~Desiati and A.~Lazarian.
\newblock Anisotropy of tev cosmic rays and the outer heliospheric boundaries.
\newblock {\em Astrophys. J.}, 762(1):44, 2013.

\bibitem{IceCubeSpectrum2013}
IceCube Collaboration.
\newblock Cosmic ray spectrum, composition, and anisotropy measured with icecube.
\newblock {\em arXiv preprint}, 2013.

\bibitem{McNally2023eleven}
F.~T. McNally and P.~Desiati et~al. (IceCube~Collaboration).
\newblock Cosmic ray anisotropy with 11 years of icecube data.
\newblock {\em arXiv preprint}, 2023.

\bibitem{IceCube2009_2015}
Silvia Bravo and IceCube Collaboration.
\newblock A closer look at the cosmic ray anisotropy with icecube.
\newblock {\em Astrophys. J.}, 2016.
\newblock Based on 2009–2015 data release.

\bibitem{IceCube_fullsky2015}
J.~C. Díaz-Vélez et~al. (IceCube \& HAWC~Collaborations).
\newblock Full‑sky analysis of cosmic‑ray anisotropy with icecube and hawc.
\newblock In {\em Proc. 34th Int. Cosmic Ray Conf.}, 2015.
\newblock arXiv:1510.04134.

\end{thebibliography}

\end{document}